\documentclass[conference]{IEEEtran}
\IEEEoverridecommandlockouts

\usepackage{cite}
\usepackage{amsmath,amssymb,amsfonts}
\usepackage{algorithmic}
\usepackage{graphicx}
\usepackage{textcomp}
\usepackage{xcolor}

\usepackage{mathptmx}  
\usepackage[T1]{fontenc}  

\usepackage{threeparttable}
\usepackage[hyphens]{url}
\usepackage{fancyhdr}
\usepackage{hyperref}
\usepackage{array}
\usepackage{booktabs}    
\usepackage{subcaption}  
\usepackage{enumitem}
\usepackage{mathtools}
\usepackage[ruled,vlined,linesnumbered]{algorithm2e} 
\usepackage{makecell}
\usepackage{tabularx}
\usepackage{tikz}
\usepackage{listings}
\usepackage{float}
\usepackage{setspace}
\usepackage{hyperref} 
\usepackage{balance}

\hypersetup{
    colorlinks=false,  
    pdfborder={0 0 0},  
    pdfstartview=FitH,
    pdftitle={FlashFuser: Expanding the Scale of Kernel Fusion},
    pdfauthor={Ziyu Huang et al.},
    pdfsubject={High Performance Computing},
    pdfkeywords={GPU, Kernel Fusion, High Performance Computing},
    pdfproducer={pdfTeX},
    pdfcreator={LaTeX with IEEEtran class}
}
\SetKwProg{Fn}{Function}{}{}
\SetKwFunction{FBestSchedule}{FindBestSchedule}
\SetKwFunction{FBestTiling}{FindBestTilingAndMapping}
\SetKwFunction{FPlaceTiles}{PlaceTiles}
\SetKwFunction{FEnumSched}{EnumerateSchedules}
\SetKwFunction{FEnumMap}{EnumerateMappings}
\SetKwFunction{FAnalyze}{CalMU} 
\SetKwFunction{Footprint}{CalMF} 
\SetKwFunction{FEval}{EvaluateTraffic}
\SetKwComment{Comment}{$\triangleright$\ }{}

\definecolor{codegreen}{rgb}{0,0.6,0}
\definecolor{codegray}{rgb}{0.5,0.5,0.5}
\definecolor{codepurple}{rgb}{0.58,0,0.82}
\definecolor{backcolour}{rgb}{0.95,0.95,0.92}
\lstdefinestyle{customcpp}{
    backgroundcolor=\color{backcolour},   
    commentstyle=\color{codegreen},
    keywordstyle=\color{magenta},
    numberstyle=\tiny\color{codegray},
    stringstyle=\color{codepurple},
    basicstyle=\ttfamily\footnotesize,
    breakatwhitespace=false,         
    breaklines=true,                 
    captionpos=b,                    
    keepspaces=true,                 
    numbers=left,                    
    numbersep=5pt,                  
    showspaces=false,                
    showstringspaces=false,
    showtabs=false,                  
    tabsize=2,
    language=C++
}
\newcolumntype{C}{>{\centering\arraybackslash}X} 
\newcommand{\proj}{FlashFuser}

\usepackage{fancyhdr} 

\fancypagestyle{firstpage}{
  \fancyhf{} 
}

\newcommand\hpcaauthors{    Ziyu Huang$^{1,2,*}$, \quad    Yangjie Zhou$^{3,*}$, \quad    Zihan Liu$^{1,2,\P}$, \quad    Xinhao Luo$^{1,2}$, \quad    Yijia Diao$^{1,2}$, \quad    Minyi Guo$^{1}$, \\ 
    Jidong Zhai$^{4}$, \quad    Yu Feng$^{1}$, \quad    Chen Zhang$^{1}$, \quad    Anbang Wu$^{1}$, \quad    Jingwen Leng$^{1,2,\P}$}
\newcommand\hpcaaffiliation{    $^1$Shanghai Jiao Tong University \qquad    $^2$Shanghai Qi Zhi Institute \\    $^3$National University of Singapore \qquad    $^4$Tsinghua University \\    \textit{$^{*}$Equal contribution \qquad $^{\P}$Corresponding authors}}
\newcommand\hpcaemail{    \{huang\_ziyu, altair.liu, lxh666, diao\_yijia, guo-my, y-feng, chenzhang.sjtu, anbang, leng-jw\}@sjtu.edu.cn, \\    yj\_zhou@nus.edu.sg, \quad    zhaijidong@tsinghua.edu.cn}

\def\BibTeX{{\rm B\kern-.05em{\sc i\kern-.025em b}\kern-.08em
    T\kern-.1667em\lower.7ex\hbox{E}\kern-.125emX}}
\begin{document}

\title{\proj{}: Expanding the Scale of Kernel Fusion for Compute-Intensive Operators via Inter-Core Connection}

\author{
    \IEEEauthorblockN{\hpcaauthors}
    \IEEEauthorblockA{\hpcaaffiliation}
    \IEEEauthorblockA{\hpcaemail}
}

\maketitle

\thispagestyle{firstpage}
\pagestyle{empty}

\begin{abstract}


The scaling of computation throughput continues to outpace improvements in memory bandwidth, making many deep learning workloads memory-bound. Kernel fusion is a key technique to alleviate this problem, but the fusion strategies of existing compilers and frameworks are limited to using local scratchpad memory. When the intermediate results exceed the limited capacity (such as FFN), the fusion fails. Although modern GPUs (like the NVIDIA H100) now incorporate an inter-core connection mechanism known as Distributed Shared Memory (DSM)—providing a larger, high-bandwidth, and low-latency on-chip memory pool—this hardware potential has yet to be exploited by software frameworks.

To bridge this gap, we present \proj{}, the first compiler framework to utilize inter-core connection for kernel fusion on modern GPUs. \proj{} extends established fusion techniques to the DSM domain through three core contributions. First, we propose a powerful DSM-based communication abstraction that formalizes complex cluster-based data exchange patterns, such as reduce, shuffle and multiply. Second, we introduce a dataflow analyzer that generalizes loop scheduling, resource mapping, and tile selection to the distributed memory hierarchy; it determines the optimal execution order and tile sizes by quantifying data movement across memory levels. Finally, \proj{} integrates these components into a unified search engine that employs analytical cost modeling and DSM-aware pruning strategies to efficiently discover the optimal execution plan. Our evaluation on an NVIDIA H100 GPU shows that \proj{} reduces memory access by 58\% and delivers kernel speedups of 3.3x against highly-tuned libraries and 4.1x against state-of-the-art compilers, resulting in a 1.24$\times$ end-to-end speedup.

\end{abstract}

\section{Introduction}

With the rapid evolution of deep learning techniques~\cite{guo2023olive, guan2024amanda, liu2024juno, guo2024gmlake, xu2024vtensor, zhou2025sample, hu2025m, zhou2025voyager, liu2025vq, guan2024fractal, zhou2023ugrapher, chen2024magis} and the expanding scale of deep learning models, the growing inference demands from multi-modal and large language models (LLM) mean that memory bandwidth is increasingly struggling to keep up with the growth of computational power.

In new generation GPU, H100, the peak FP16 compute capability has increased to approximately 1000\,TFLOPS from the 300\,TFLOPS of the previous-generation A100 (3.3$\times$ increase), while the global HBM bandwidth has only grown from 2\,TB/s to 3\,TB/s (1.5$\times$ increase)\cite{choquette2020nvidia,choquette2022nvidia, luo2024benchmarking, abdelkhalik2022demystifying}. This disparity, known as the memory wall, in growth rates makes memory bandwidth a primary bottleneck. In workloads dominated by General Matrix Multiplication (GEMM), such as Transformer Feed Forward Network (FFN) layers and convolutional blocks, insufficient HBM bandwidth often becomes a significant bottleneck. As shown in Table~\ref{tab:ffn_percentage}, under a typical inference configuration with a sequence length of 512, the FFN in various models consumes 40\%--60\% of the total execution time\cite{wei2024building} and exhibits memory-bound characteristics. 



\begin{table}[t]
    \centering
    \caption{Percentage of Execution Time Spent in FFN Layers across Different Models.}
    \label{tab:ffn_percentage}
    \begin{tabular}{l c} 
        \toprule
        \textbf{Model} & \textbf{FFN Time (\%)} \\
        \midrule
        GPT-6.7B    & 61.28 \\
        LLaMA-1B    & 57.44 \\
        OPT-1.3B    & 53.08 \\
        BERT        & 47.03 \\
        GPT-2       & 41.64 \\
        \bottomrule
    \end{tabular}
\end{table}


To mitigate the aforementioned bandwidth bottleneck, modern GPUs such as the H100 have introduced inter-core connected architecture, which provides a high-speed data exchange path known as Distributed Shared Memory (DSM) within a cluster composed of multiple Streaming Multiprocessors (SMs)\cite{NVIDIA_CUDA_DSM_2025}. The traditional approach relies on a costly round-trip path through global memory, whereas our approach leverages DSM to open up a direct on-chip path. This shift in the data path has two benefits. First, by avoiding the redundant ``write-then-read'' operation, the total volume of data transferred to and from global memory is significantly reduced. Second, this direct on-chip path provides both higher bandwidth and lower latency than global memory access\cite{luo2024benchmarking}.


Kernel fusion is an effective method for addressing the aforementioned memory-bound problem. However, current kernel fusion techniques fail to fuse large-scale operator chains. Existing software frameworks—including libraries like \texttt{cuBLAS}~\cite{NVIDIA_cuBLAS_2025} and \texttt{CUTLASS}~\cite{Thakkar_CUTLASS_2023}, inference frameworks like \texttt{SGLang}~\cite{zheng2024sglang}, or compilers like Chimera~\cite{zheng2023chimera}, BOLT\cite{xing2022bolt}—typically handle smaller operator chains by placing intermediate results in the shared memory (SMEM) or registers(reg) of a single SM. When the intermediate data becomes larger (like FFN), these methods will abandon fusion, resorting to an inefficient round-trip to global memory. The inter-core connection mechanism can effectively alleviate this constraint. By interconnecting the SMEM of multiple SMs, it creates what can be viewed as an expanded on-chip memory pool. However, the complex communication patterns required to leverage this capability remains an unexplored domain.


To bridge the gap between new hardware features and existing software frameworks, we propose \proj{}, the first deep learning (DL) compiler to leverage DSM for kernel fusion on compute-intensive operator chains. By creatively introducing DSM, \proj{} expands the scope of fusible operators. It progressively places intermediate results to on-chip memory, including reg, SMEM, and DSM, thereby introducing a vast search space. Through corresponding pruning rules and cost model, it finds an execution order that minimizes data movement, thus achieving a performance improvement.



\begin{figure}[t]
    \centering 
    \begin{minipage}[b]{0.48\linewidth}
        \centering
        \includegraphics[width=\linewidth]{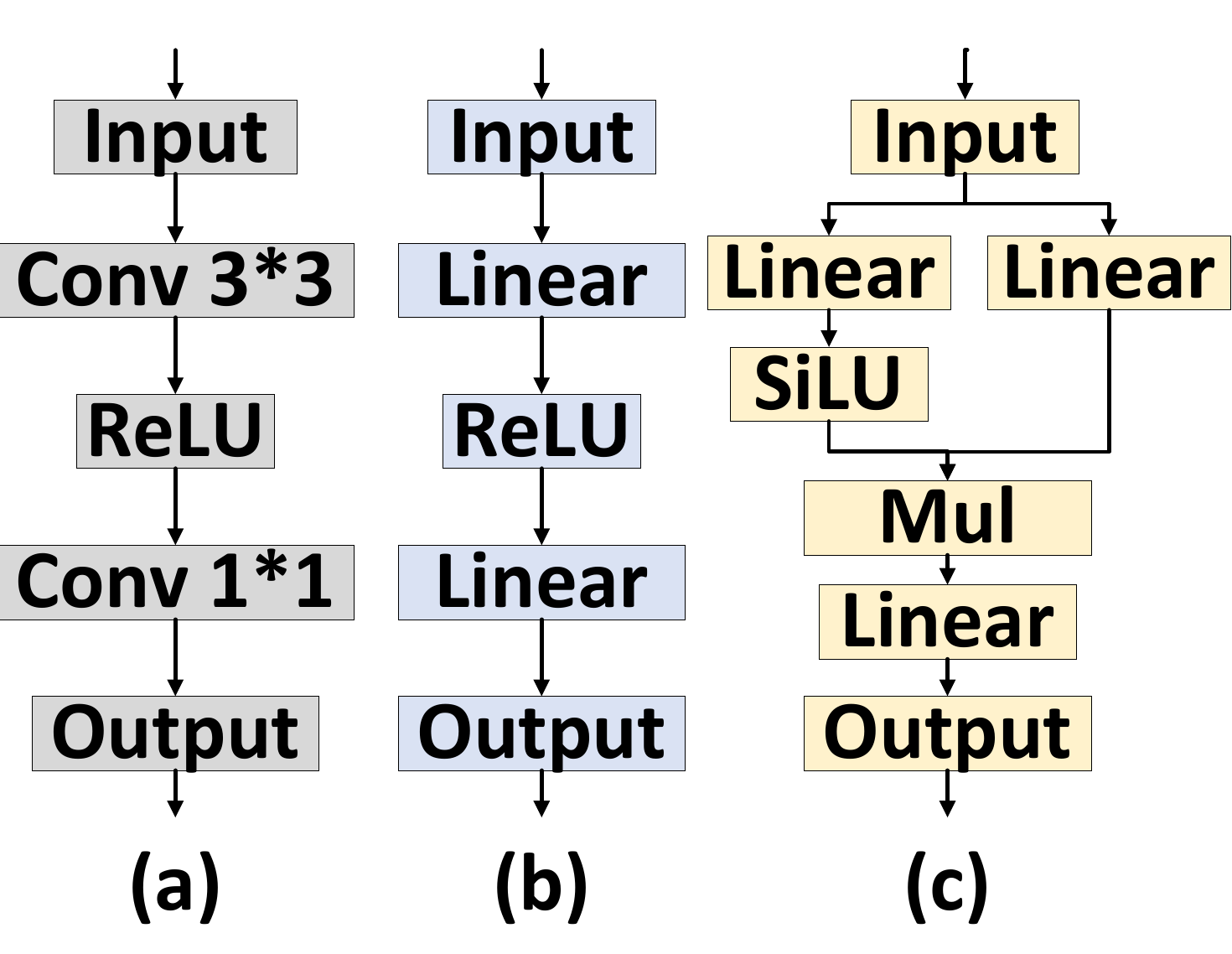}
        \captionof{figure}{Three common GEMM chains: (a) conv, (b) standard FFN, (c) gated FFN.}
        \label{fig:background_cases}
    \end{minipage}
    \hfill 
    \begin{minipage}[b]{0.48\linewidth}
        \centering
        \includegraphics[width=\linewidth]{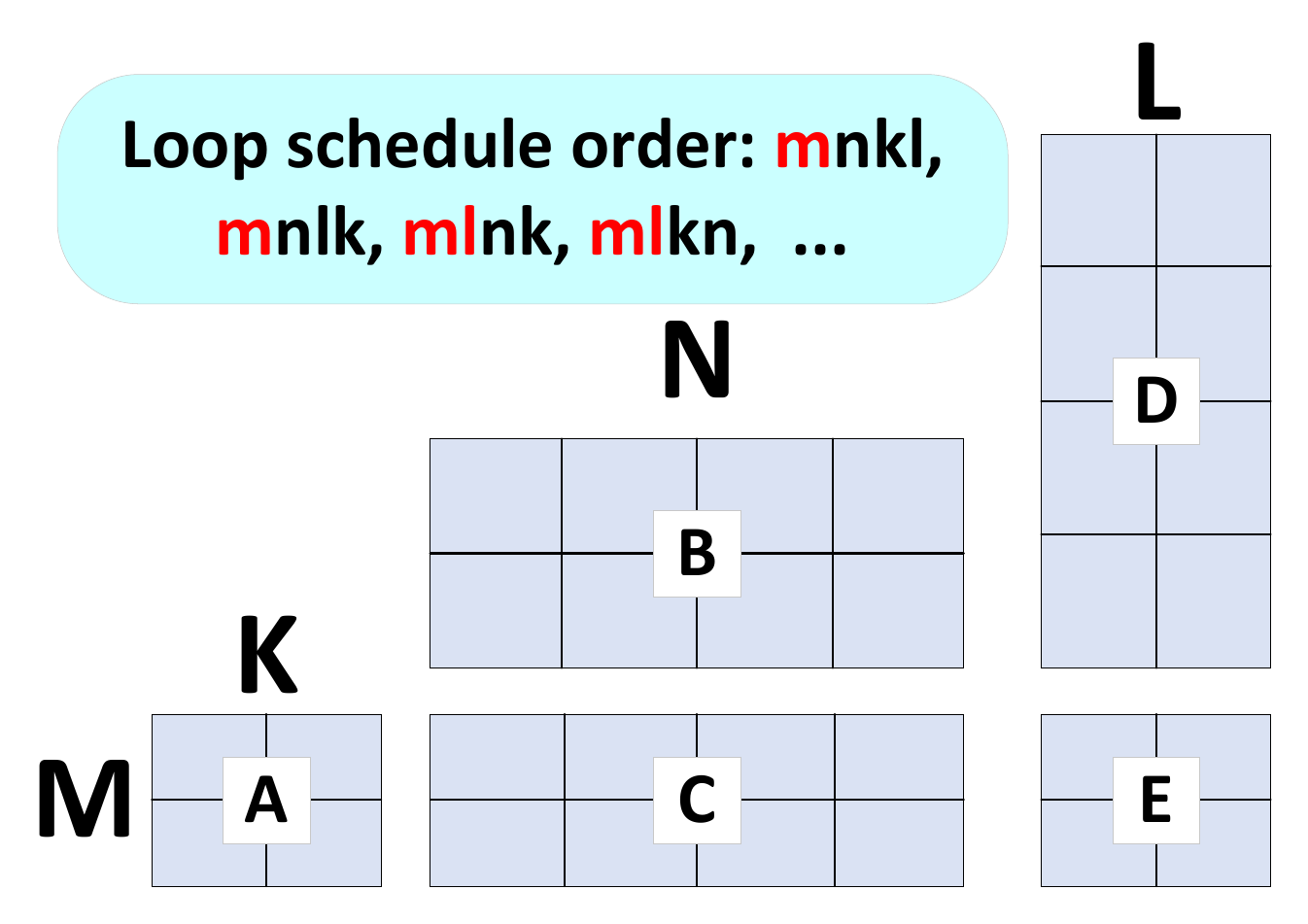}
        \captionof{figure}{A fused GEMM operator chain, showing its loop dimensions (M, N, K, L) and possible execution orders (mnkl, etc.).}
        \label{fig:MNKT_def}
    \end{minipage}
\end{figure}


We use compute intensive operator chains from various LLM and CNNs for performance evaluation, running on an NVIDIA H100 GPU. \proj{} achieves a speedup of up to 4.1x over the state-of-the-art (SOTA) baseline. In summary, the contributions of this paper are as follows:

\begin{itemize}[leftmargin=*]
    \item We identify the operator fusion bottleneck caused by SMEM capacity limitations, and point out the widespread deficiency in the current software ecosystem in utilizing inter-core connection property (\S\ref{sec:motivation}). 
    \item We propose a new abstraction to describe inter-core communication patterns, enabling it to support the various requirements of kernel fusion (\S\ref{ssec:dsm_comm_primitive}).
    \item We propose a dataflow analyzer that quantifies data-movement cost across the memory hierarchy and schedules data to spill progressively from fast to slow caches (\S\ref{ssec:Spilling}).
    \item We present a fusion search engine that employs pruning techniques and an analytical cost model to efficiently navigate the greatly expanded search space introduced by DSM (\S\ref{ssec:search_engine}).
    \item Compared to highly-tuned libraries and state-of-the-art compilers, our method delivers kernel speedups of 3.3$\times$ and 4.1$\times$, respectively, along with a 58\% reduction in memory access. These kernel-level improvements result in a 1.24$\times$ end-to-end speedup, validating the effectiveness of our approach (\S\ref{sec:evaluation}).
\end{itemize}

\section{Background}



Mainstream LLM and Convolutional Neural Networks (CNNs) consist of numerous tensor operators, which are often organized into chains. As shown in Figure~\ref{fig:background_cases}, these include convolution blocks that can be converted to GEMM chains via \texttt{im2col}~(a), standard FFN~(b), and Gated FFNs with branched structures (e.g., SwiGLU)~(c). Due to their data-intensive nature, these GEMM-based operator chains are often limited by memory bandwidth, which makes kernel fusion a key optimization method. Figure~\ref{fig:MNKT_def} shows an example of a GEMM chain, where the dimensions of each matrix are marked as M, N, K, and L. In parallel computing, we need to split the tensors into small blocks and then iterate through these blocks according to different iteration orders. This traversal order is called a loop schedule, and as shown in the Figure~\ref{fig:MNKT_def}, can be \texttt{mnkl}, \texttt{mnlk}, etc.

\begin{figure}[t]
    \centering
    \includegraphics[width=\linewidth, height=0.8\textheight, keepaspectratio]{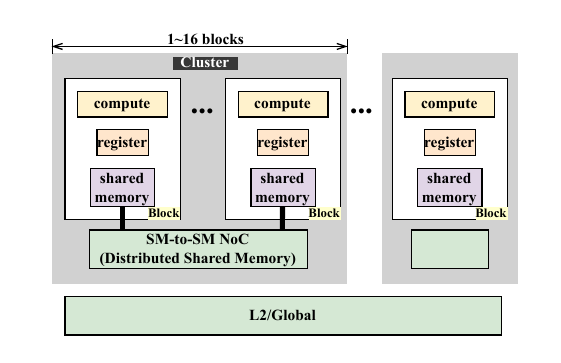}
    \caption{The memory hierarchy of the H100 GPU, including registers, SMEM, DSM, L2 cache, and global memory.}
    \label{fig:dsm_struct}
\end{figure}


DSM is one tier in the multi-level cache hierarchy. Figure~\ref{fig:dsm_struct} illustrates the entire cache hierarchy of the H100 GPU. The innermost cache is the L0 cache\cite{riedel2023mempool}, also known as the register file (reg). This is the fastest cache, but it is only visible to each thread and has a small capacity. The next tier is the L1 cache, or SMEM\cite{nvidia_cuda_guide}, where all threads within a single compute core can access the values in SMEM. Starting from the H100, the SMEMs of different SMs can be connected via DSM, which is also considered an L1.5 cache\cite{falahati2024cross, ibrahim2020analyzing, ibrahim2021analyzing}. Only cores within a single cluster can exchange data, different clusters cannot directly interact and must exchange data through the next tier, the L2 cache and the global memory.

  
  

\begin{figure}[t]
    \centering
    \includegraphics[width=\linewidth, height=0.8\textheight, keepaspectratio]{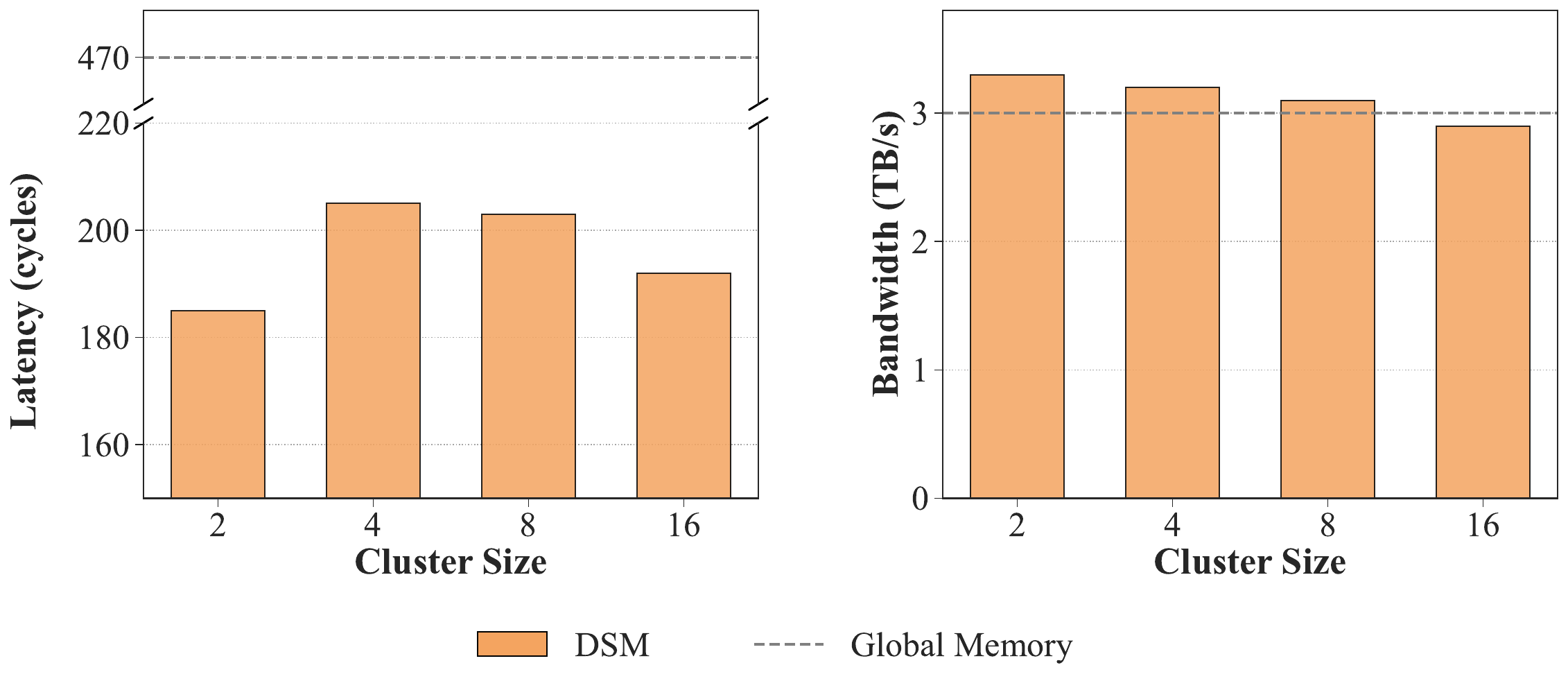}
    \caption{Bandwidth and latency of DSM under different cluster sizes. The corresponding performance of global memory is marked in the figure for comparison.}
    \label{fig:dsm_perf}
\end{figure}

\section{Motivation} \label{sec:motivation}

\begin{figure}[b]
    \centering
    \includegraphics[width=\linewidth, height=0.8\textheight, keepaspectratio]{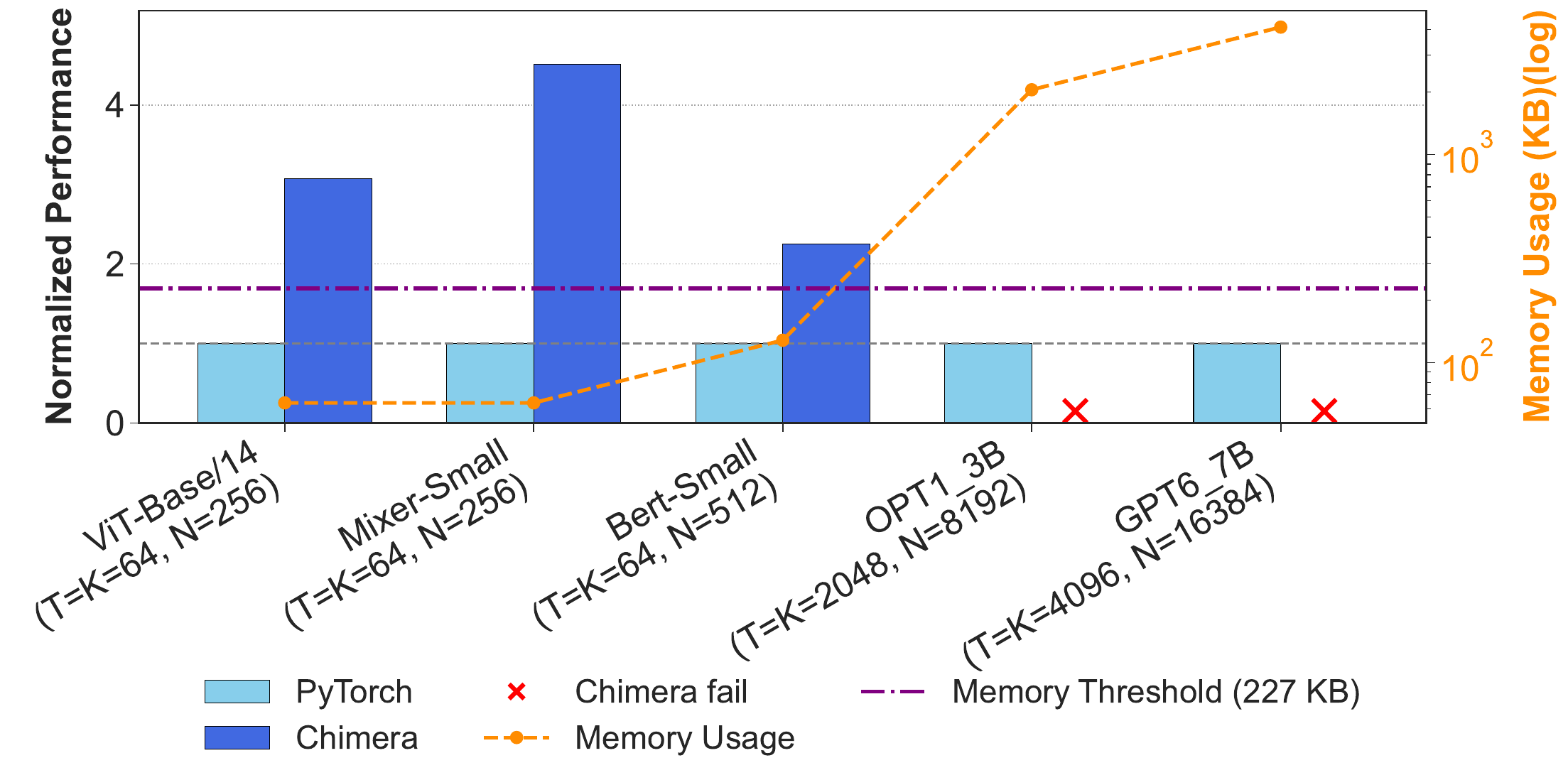}
    \caption{Relative performance of Chimera to torch. The workload consists of two consecutive GEMM operations. M is set as 128 here.}
    \label{fig:chimera_fail}
\end{figure}

The gap between new hardware features and the software ecosystem is characterized by two core limitations in existing works:

\textit{\textbf{(a) Fusion is constrained by on-chip memory capacity:}} Current kernel fusion frameworks are constrained by SMEM capacity, which prevents the fusion of large-scale operator chains. Current frameworks only consider reg and SMEM for data reuse and make the overly simplistic assumption that intermediate results can always be accommodated on-chip; however, this assumption does not hold true in many scenarios. As illustrated in Figure~\ref{fig:chimera_fail}, while Chimera\cite{zheng2023chimera} can still store intermediate results on-chip when memory usage is relatively small, it encounters fusion failures when executing larger-scale GEMM chains, such as those in OPT1\_3B and GPT6\_7B. As shown by the purple dotted line in the figure, the upper limit of SMEM for a single SM on H100 is 227KB. When the intermediate result exceeds this size, the fusion will fail.

\begin{table}[t]
    \centering
    \small 
    \caption{Comparison of \proj{} with representative previous works. Cache Hierarchy 0,1,1.5 means register(reg), shared memory(SMEM) and dsm}
    \label{tab:comparison}
    
    \renewcommand{\arraystretch}{0.9}
    \setlength{\tabcolsep}{3pt} 
    
    \resizebox{\columnwidth}{!}{%
        \begin{tabular}{l l l c c} 
            \toprule
            \textbf{Framework} & \textbf{Cache Hier.} & \textbf{Strategy} & \textbf{GPU Supp.} & \textbf{Fusion} \\
            \midrule
            BOLT\cite{xing2022bolt}      & 0/1       & Tuning      & yes & yes \\
            Chimera\cite{zheng2023chimera} & 1         & Analytical  & yes & yes \\
            Welder\cite{shi2023welder}     & 0/1       & Analytical  & yes & yes \\
            MCFuser\cite{zhang2024mcfuser} & 1         & Analytical  & yes & yes \\
            T10\cite{liu2024scaling}       & 1/1.5     & Analytical  & no  & no \\
            WaferLLM\cite{he2025waferllm}  & 1/1.5     & Handcrafted & no  & no \\
            \midrule 
            \textbf{Ours}          & \textbf{0/1/1.5} & \textbf{Analytical} & \textbf{yes} & \textbf{yes} \\
            \bottomrule
        \end{tabular}%
    }
\end{table}

\textit{\textbf{(b) DSM can expand on-chip memory, however, its performance is non-trivial:}}
By connecting multiple SMs, DSM provides an effectively larger SMEM space, which can solve the problem of SMEM size limitation. However, its complex characteristics make it difficult to utilize directly. \textbf{Firstly,} the bandwidth and latency of DSM vary with cluster size. As shown in Figure~\ref{fig:dsm_perf}, as the cluster size increases, its latency tends to increase while its bandwidth gradually decreases. For all cluster sizes, the DSM latency is lower than that of global memory, and for all but the largest cluster size, its bandwidth is faster~\cite{luo2024benchmarking, luhnen2024benchmarking, jin2024uncovering}, making the selection of an appropriate cluster size a non-trivial problem. \textbf{Secondly,} the introduction of clusters adds another layer to the memory hierarchy, making dataflow more complex. Since prior works did not incorporate DSM, they are unable to analyze how data should be placed on DSM or the resulting data movement volume across various cache levels. This analysis involves crucial details such as how tiles are partitioned, their execution order, their sizes, and resource mapping. \textbf{Finally,} the introduction of DSM makes many previously infeasible fusion scenarios possible. This is because considering DSM is equivalent to expanding the on-chip memory space, making more strategies feasible that would have been directly pruned in prior works. As detailed in Section~\ref{ssec:search_engine}, for GPT6\_7B, the number of possibilities with traditional methods\cite{zhang2024mcfuser} after pruning is approximately $10^4$, whereas with DSM, this expands to $10^6$. Therefore, an analysis framework specifically targeted for DSM is essential.

Prior works, as summarized in Table~\ref{tab:comparison}, have only partially addressed the aforementioned issues. Chimera\cite{zheng2023chimera} and MCFuser\cite{zhang2024mcfuser} considered how to fuse GEMM chains, but because they only use SMEM to store intermediate results, they are severely limited by the SMEM capacity of a single SM and thus cannot be used for scenarios with larger intermediates, such as FFNs. BOLT\cite{xing2022bolt} considered how to use registers or SMEM to perform GEMM chain fusion; however, it did not consider different computation orders and used manual tuning to find parameters, meaning its search results are not necessarily optimal. Welder\cite{shi2023welder} used an analytical method to explore data reuse for reg and SMEM, but it also did not consider DSM. Previous papers on DSM, such as T10\cite{liu2024scaling} and WaferLLM\cite{he2025waferllm}, are works on Graphcore and Cerebras, respectively; both utilized inter-core connect features but did not consider kernel fusion, nor were they explored on GPUs.

\section{Design} \label{sec:design}

\begin{figure}[t]
    \centering
    \includegraphics[width=\linewidth, height=\textheight, keepaspectratio]{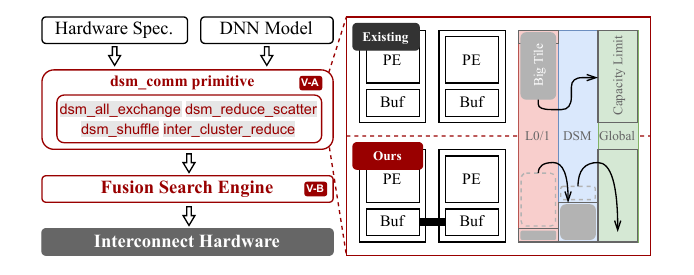}
    \caption{System overview of \proj{}}
    \label{fig:system_design_new}
\end{figure}

We now introduce \proj{}, a compiler designed to optimize kernel fusion for operator chains on processors with inter-core connection. An overview of \proj{} is presented in Figure~\ref{fig:system_design_new}:

(1) \proj{} defines a DSM-communication primitive that compactly encodes SM partitioning and inter-SM dataflow, yielding a unified representation of DSM-based fusion plans under the given model and hardware (see \S\ref{ssec:dsm_comm_primitive}).

(2) Based on this representation, our dataflow analyzer evaluates the feasibility and cost—in terms of data movement volume—of any given plan. It models the entire on-chip memory hierarchy, determining how intermediate data is progressively spilled from high-speed caches to slower tiers (like DSM) when capacity is exceeded (see \S\ref{ssec:Spilling}).

(3) The incorporation of DSM unlocks many new fusion possibilities, thereby creating an enormous search space. To navigate this, \proj{} employs a fusion search engine. Guided by a cost model and a set of pruning rules, the engine efficiently searches for the optimal execution plan (see \S\ref{ssec:search_engine}).

Methodologically, \proj{} adapts established techniques from prior kernel fusion works, including loop scheduling, tile selection, and resource mapping (see \S\ref{ssec:Spilling}), as well as cost modeling and pruning (see \S\ref{ssec:search_engine}). \textbf{However, our fundamental novelty lies in integrating DSM into these methods}---specifically, by introducing DSM-level tiling, accounting for DSM bandwidth variations across cluster sizes, and respecting maximum cluster size limits, etc.

\subsection{dsm\_comm primitive}
\label{ssec:dsm_comm_primitive}

\begin{figure*}[t]
    \centering
    \includegraphics[width=\textwidth]{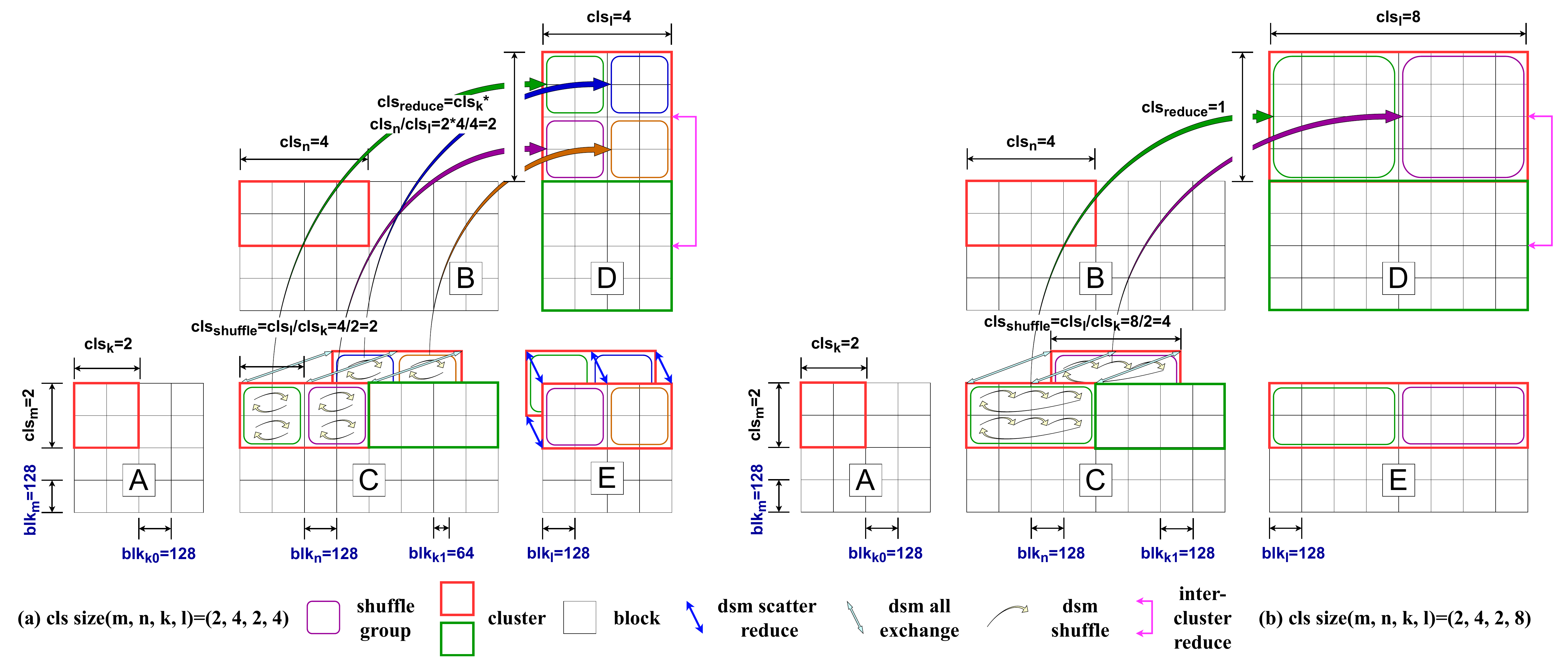}
    \caption{Conceptual illustration of the cluster and tile geometries.}
    \label{fig:cluster_tile_concept}
\end{figure*}

Conventional GPU programming models have primarily focused on a single tiling hierarchy at the thread block level. The introduction of DSM necessitates a higher-level tiling at the thread block cluster level, which in turn requires explicit handling of intra-cluster and inter-cluster communication.

To elaborate on this design, we use a fused kernel containing two GEMM operations as an example. The execution of this fused kernel is divided into three distinct phases: $GEMM_0$, $GEMM_1$, and Store phase (as illustrated in Figure~\ref{fig:cluster_tile_concept}). In this context, a bold rectangle denotes a Cluster, a non-bold rectangle denotes a Block, and a rounded rectangle represents a Shuffle Group, where Blocks within it perform shuffle operations to exchange data. We define two base parameters: $cls_i$, representing the number of parallel Blocks within a Cluster along dimension \textit{i}, and $blk_i$, representing the data granularity computed by a Block along dimension \textit{i}.

\textbf{Crucially, the dataflow between blocks in the GEMM chain is uniquely determined by the declared cluster size.} In the two-GEMM scenario, these dimensions correspond to $cls_m, cls_n, cls_k, \text{and } cls_l$. As shown in Figure~\ref{fig:cluster_tile_concept}(a), the cluster size is (2, 4, 2, 4). In the $GEMM_0$ phase, $cls_k=2$ signifies that the K-dimension is spatially partitioned across two parallel Blocks. Consequently, these Blocks must perform an intra-cluster accumulation along the K-dimension. We introduce the \texttt{dsm\_all\_exchange} primitive for this purpose, ensuring each Block holds the complete, fully-accumulated intermediate result before proceeding.

In the $GEMM_1$ and Store phases, data must be shared among the Blocks to compute the final matrix E. We employ two complementary strategies—shuffle and reduce—to compute matrix E. The first strategy is a shuffle, where data is exchanged during the $GEMM_1$ computation. As shown in Figure~\ref{fig:cluster_tile_concept}(a), calculating one Block of matrix E requires access to an entire row of data from matrix C. Therefore, Blocks within the same Shuffle Group exchange their respective slices of matrix C using the \texttt{dsm\_shuffle} primitive.

The second strategy is reduce, which postpones data exchange to the final Store phase. Here, each Block first independently computes a partial sum of the output matrix E. This is followed by a two-level hierarchical reduction. The intra-cluster reduction is performed first, where multiple contributing Shuffle Groups perform an accumulation via the \texttt{dsm\_reduce\_scatter} operation. The Scatter pattern is employed because each Block is only responsible for writing back a portion of the final result, thus avoiding data redundancy. This is followed by the \texttt{inter\_cluster\_reduce}, which aggregates partial sums from all participating clusters. This step is implemented by leveraging the NVIDIA Hopper architecture's Tensor Memory Accelerator (TMA). Through its \texttt{cp.reduce.async.bulk} instruction, the TMA can asynchronously perform atomic reductions across clusters. 

To precisely describe these communication patterns, we derive two key variables based on the established cluster parameters: $cls_{\text{shuffle}}$, the number of Blocks within a single Shuffle Group, and $cls_{\text{reduce}}$, the number of Shuffle Groups participating in a Reduce operation. Their derivations are $cls_{\text{shuffle}} = cls_l / cls_k$ and $cls_{\text{reduce}} = cls_n / cls_{\text{shuffle}} = (cls_n \times cls_k) / cls_l$. For instance, Figure~\ref{fig:cluster_tile_concept}(b) illustrates an alternative configuration where the \texttt{cls size} is (2, 4, 2, 8). Here, $cls_{\text{reduce}}=1$, meaning no inter-group reduction is needed during the Store phase. The trade-off is that the larger $cls_{\text{shuffle}}$ increases the communication volume for the shuffle operations, while decreasing the number of required \texttt{dsm\_reduce\_scatter} operations. Moreover, this flexibility to configure the shuffle and reduce dimensions is crucial for efficiently mapping problem sizes that are small or not perfectly divisible onto the hardware.

\begin{figure}[t]
    \centering
    \includegraphics[width=\linewidth, height=\textheight, keepaspectratio]{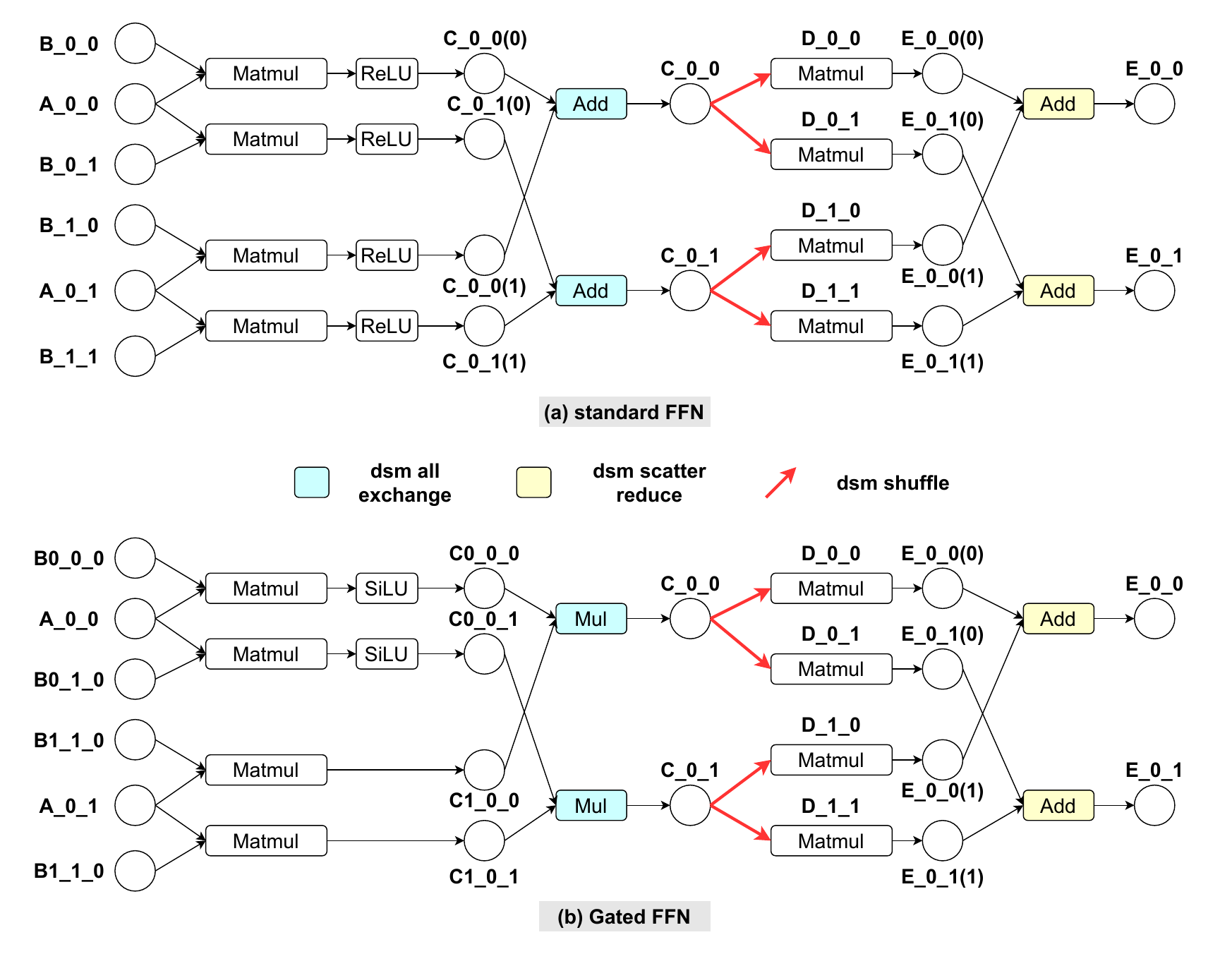}
    \caption{Tile graph of kernel fused with the \texttt{dsm\_comm} primitive (only show one cluster). (a) standard FFN. (b) gated FFN}
    \label{fig:tile_graph}
\end{figure}

Based on the aforementioned DSM communication primitives, we can abstract complex fused kernels into an intuitive tile graph to describe the dataflow. As shown in Figure~\ref{fig:tile_graph}, the graph clearly demonstrates the flexibility of our framework: it can support not only standard FFNs composed of consecutive GEMMs (Figure~\ref{fig:tile_graph}(a)) but also the more structurally complex Gated FFN variant (Figure~\ref{fig:tile_graph}(b)).

We use the standard FFN in Figure~\ref{fig:tile_graph}(a) to illustrate the detailed dataflow. A symbol such as $B_{0,0}$ denotes the tile of matrix B at coordinate (0,0). The process begins with tiles like $A_{0,0}$ and $B_{0,0}$ being multiplied to produce a partial sum of the intermediate matrix C, denoted as $C_{0,0}(0)$. After all parallel partial sums (e.g., $C_{0,0}(0)$ and $C_{0,0}(1)$) are computed, the \texttt{dsm\_all\_exchange} primitive performs an All-Reduce operation within the cluster to produce the complete intermediate tile $C_{0,0}$. Subsequently, $C_{0,0}$ serves as input to GEMM1 and is distributed by the \texttt{dsm\_shuffle} primitive (red arrows) to different compute units to be multiplied with different tiles of matrix D, yielding new partial sums for matrix E (e.g., $E_{0,0}(0)$ and $E_{0,1}(0)$). Finally, during the Store phase, these partial sums of E are accumulated by the \texttt{dsm\_scatter\_reduce} primitive to obtain the complete output tiles $E_{0,0}$ and $E_{0,1}$.

For the Gated FFN in Figure~\ref{fig:tile_graph}(b), the core difference is that its Up-FFN portion executes two parallel GEMM branches, where the result of one branch, after a SiLU activation, is element-wise multiplied with the other. This impacts the function of the first DSM primitive: \texttt{dsm\_all\_exchange} now performs a \texttt{Mul} operation instead of an \texttt{Add}, which is why we chose the generic name ``exchange'' to reflect its operational flexibility. To implement the two parallel GEMM branches, our framework supports two approaches. The first is to leverage spatial partitioning by setting \texttt{cls\_k = 2}, which assigns the two GEMM branches to different groups of Blocks. The final element-wise multiplication is then performed by the \texttt{dsm\_all\_exchange} primitive, which executes a \texttt{Mul} operation to combine the results. The second approach is to execute the two GEMMs sequentially within a single Block, which effectively transforms the computation into the pattern of a standard FFN, but with a doubled K-dimension. 

The two approaches allow for optimizing different goals: the first, spatial partitioning across the cluster, is designed to maximize parallelism, while the second, sequential execution within each Block, aims to minimize DSM communication overhead.

\subsection{Dataflow Analyzer} \label{ssec:Spilling}

\begin{algorithm}[htbp]
\caption{Dataflow Analyzer}
\label{alg:spilling_planner}

\begin{spacing}{0.9} 

\SetAlgoSkip{smallskip} 

\SetKwInOut{Input}{Input}
\SetKwInOut{Output}{Output}
\Input{
    Graph $g$, Device $d$, \\
    Loop schedule $s = \{s_1, s_2, \dots, s_x\}$, \\
    Tile sizes $t = \{t_1, t_2, \dots, t_x\}$,\\ 
    Initial resource mapping $r$
}
\Output{
    Data movement volume $D_V$, Final plan $p_{final}$
}

\SetKwFunction{FMain}{DataflowAnalyzer}
\SetKwFunction{FGetFootprint}{GetFootprint}
\SetKwProg{Fn}{Function}{:}{}
\Fn{\FMain{$g, d, s, t, r$}}{
    $mapping\_plan \leftarrow \text{new ResourceMapping()}$\;
    $D_V \leftarrow 0$\;
    $hierarchy \leftarrow d.\text{getMemoryHierarchy()}$\;
    $S \leftarrow g.\text{getDimensionSizes()}$\;
    
        \ForEach{tensor in $g.tensors()$}{
            $DF \leftarrow \FGetFootprint(t.block)$
            
            \uIf{$tensor \in g.IOTensors()$}{
                $DM \leftarrow DF$\;
                \ForEach{$s_i$ in reversed($s$)}{
                    \If{$s_i$ accesses $tensor$}{
                        $DM \leftarrow DM \times \lceil S_i / t_i.block \rceil$
                    }
                }
                $D_V[global] \leftarrow D_V[global] + DM$\;
            }
            \Else{
                $remaining \leftarrow DF$\;
                $mapping \leftarrow \text{new TensorMapping()}$\;
                \tcp{Greedily place tensor across memory hierarchy}
                \ForEach{level in $hierarchy$}{
                    \If{$remaining \le 0$}{
                        \nl\KwSty{break}\;
                    }
                    $alloc \leftarrow \min(remaining, level.capacity)$\;
                    $mapping[level] \leftarrow alloc$\;
                    $remaining \leftarrow remaining - alloc$\;
                    \ForEach{$s_i$ in reversed($s$)}{
                        \If{$s_i$ accesses $tensor$}{
                            $D_V[\text{level}] \leftarrow update\_dv (t_i.cluster, DF);$
                        }
                    }
                }
                $mapping\_plan[tensor] \leftarrow mapping$\;
            }
        }
    
    $p_{final} \leftarrow (s, t, mapping\_plan)$
    
    \KwRet{$(D_V, p_{final})$}\;
}
\end{spacing}
\end{algorithm}

After introducing the \texttt{dsm\_comm} primitive, we incorporate it into our \textbf{Dataflow Analyzer}. This analyzer is designed to tackle the complex dataflow challenges introduced by inter-core connection. While conventional methods only need to consider the register and SMEM hierarchy, our approach must also orchestrate the newly introduced DSM tier. To address this, \proj{} employs a tile-based analysis method. For a given set of parameters, the analyzer determines how to efficiently place intermediate data for reuse across the memory levels. Crucially, it also analyzes the data movement in detail, allowing it to calculate critical performance costs, such as the data transfer volume for each tier of the memory hierarchy. 


\subsubsection{Loop Scheduling}
\label{Loop_Schedule}
The \texttt{LoopSchedule} defines the loop execution order for a operator chain. First, we unify the codependent loop dimensions from all operators into a single independent set, formally denoted as $\mathcal{X} = \{x_0,x_1,\dots,x_{J-1}\}$. This set is then scheduled by defining a permutation $s$ to set the nesting order and partitioning the dimensions into \emph{spatial} ($\mathcal{S}$) or \emph{temporal} ($\mathcal{T}$). \emph{Spatial} refers to using multiple parallel processing units to compute a dimension simultaneously, while \emph{temporal} refers to using a single processing unit to sequentially compute an entire dimension over time.

\begin{figure}[t]
    \centering
    \includegraphics[width=\linewidth, height=0.8\textheight, keepaspectratio]{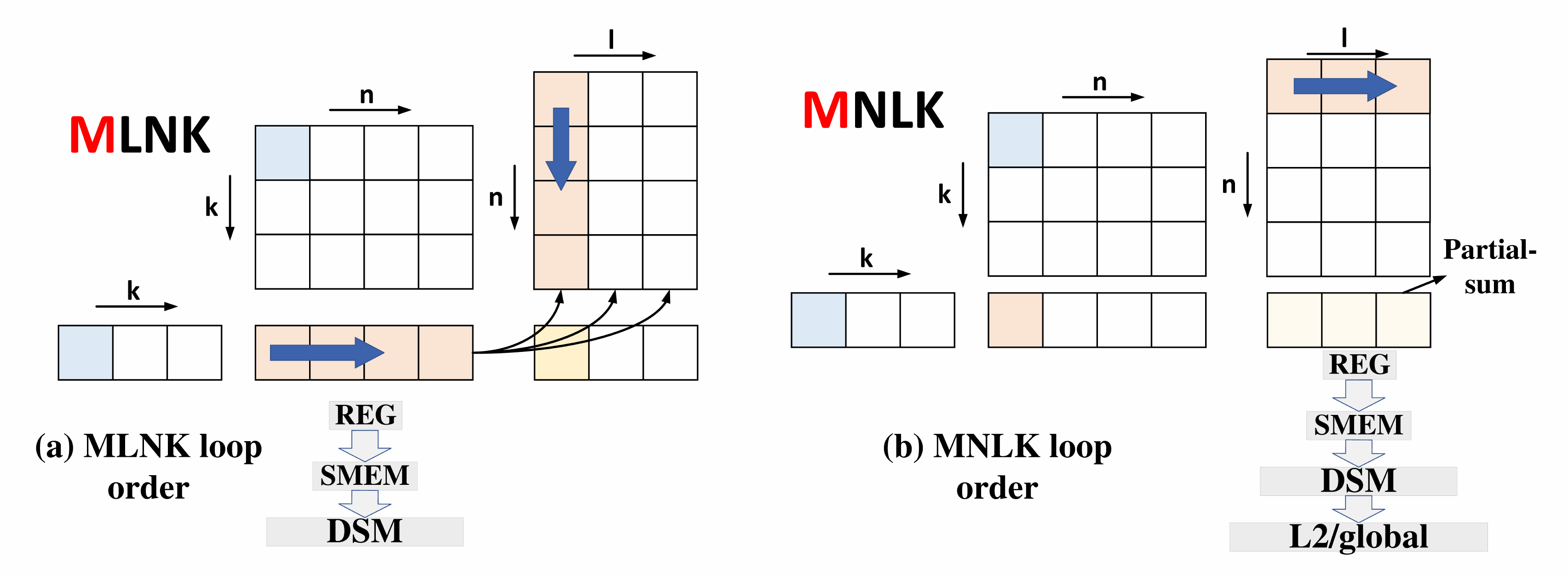}
    \caption{An example of the hierarchical spilling plan, illustrating different spilling strategies. The red `M' denotes the spatial dimension, while the black letters represent temporal dimensions.}
    \label{fig:spill_example}
\end{figure}

Different loop schedules affect the size of the tensor that needs to be cached. As illustrated in Figure~\ref{fig:spill_example}, the MLNK order in (a) requires the local block to store the complete tensor C. Depending on the hardware speculation and problem size, this may require spilling from reg to SMEM, or further to DSM. In contrast, the MNLK order in (b) produces a partial E after each iteration of the LK loops. Although accumulating in registers is most efficient, the limited register space may necessitate spilling to SMEM, DSM, or even L2/global.




\subsubsection{Tile Selection}

The tile size is defined across three hierarchical levels: a cluster-level vector (\texttt{tile.cluster}) that dictates how work is distributed across clusters, a block-level vector (\texttt{tile.block}) that governs the tile size computed by each block.

This tiling directly impacts memory usage and dataflow patterns. The block-level factors (\texttt{tile.block}) determine the data tile size each thread block must hold, influencing the decision of whether to use registers or shared memory. The cluster-level factors (\texttt{tile.cluster}) influence data distribution across SMs, thereby determining whether intermediate data must spill to global memory and dictating the inter-block data exchange patterns.

\subsubsection{Resource Mapping}
\label{Resource_Mapping}
Our framework binds tensors to different memory tiers through a heuristic-driven approach. This process, detailed in Algorithm~\ref{alg:spilling_planner}, analyzes the memory usage of reusable tensor —as determined by the \texttt{LoopSchedule} and \texttt{TilingSize}—and then generate its data reuse plan across the cache hierarchy.

We use the \texttt{DataflowAnalyzer} to generate a concrete spilling plan for reused tensor. This function takes a computation graph ($g$), device information ($d$), a loop schedule ($s$), a tile size ($t$), and an initial resource mapping ($r$) as its inputs. 

First, we obtain the size of each dimension (M, N, K, L) from the graph. If a dimension is spatial, a full traversal is not required, so its effective size is set to the corresponding tile size (line 5). We use the `getFootprint' function to obtain the data access volume within a single tile (line 7). For input and output tensors, we calculate their total data movement volume from global memory. This is achieved by iterating through all dimensions; for each dimension relevant to the tensor, the data movement is multiplied by a factor that accounts for the increased accesses caused by tiling (lines 8-13). For a reused tensor, it is not necessarily placed in a single memory level; it can be distributed across multiple levels. Its data footprint (DF) determines the required memory size. A greedy algorithm is then employed to place the tensor on the highest-level memory possible. If a level's capacity is exceeded, the remaining portion is spilled to the subsequent level in the hierarchy (lines 17-23). Throughout this process, the data movement volume for each cache level is calculated. Since DSM has a lower bandwidth than SMEM, our analysis primarily focuses on the DSM traffic. As described in \texttt{dsm\_comm}, we calculate the DSM traffic for either Standard FFN or Gated FFN based on the cluster size and data footprint, thereby deriving the data movement volume(lines 23-26). Finally, the algorithm outputs the total data movement volume and the final plan, which consists of the determined resource mapping, together with loop schedule and tile size (lines 27-29).



\subsection{Fusion Search Engine}\label{ssec:search_engine}

Our search engine is designed to efficiently explore the vast search space composed of loop schedules, tiling sizes, and resource mapping to find the optimal fusion plan. Its core principle is to leverage an analytical cost model and pruning strategies to rapidly filter out a large number of inefficient or incorrect candidates.

\subsubsection{Cost Model} \label{ssec:cost_model}

Our performance model is inspired by the analytical model in Chimera \cite{zheng2023chimera}. We model the data movement cost across the $L$ levels of the memory hierarchy. The cost $C_l$ of transferring data to level $l$ is determined by the required data volume $V_l$ for a given tiling strategy $\mathcal{T}_l$, and the memory bandwidth $B_l$ of that level.

\begin{equation}
C_l(\mathcal{T}_l) = \frac{V_l(\mathcal{T}_l)}{B_l} \tag{1}
\end{equation}

To optimize the overall performance, we aim to minimize the bottleneck, which is the slowest data movement stage among all memory levels. This is formulated as a minimax optimization problem:

\begin{equation}
\min_{\mathcal{T}_1, \dots, \mathcal{T}_L} \left\{ \max_{l=1, \dots, L} \left( C_l(\mathcal{T}_l) \right) \right\} \tag{2}
\end{equation}

The optimization is subject to memory capacity constraints of each level, where the memory usage $U_l$ dictated by the tiling strategy $\mathcal{T}_l$ cannot exceed the available capacity $\text{Cap}_l$.

\begin{equation}
\text{s.t.} \quad U_l(\mathcal{T}_l) \le \text{Cap}_l, \quad \forall l \in \{1, \dots, L\} \tag{3}
\end{equation}

\subsubsection{Pruning Strategies} \label{ssec:pruning}

\begin{table}[t]
\centering
\caption{Pruning results based on rules}
\label{tab:pruning_effectiveness}
\small 
\setlength{\tabcolsep}{4pt} 
\renewcommand{\arraystretch}{0.9} 
\begin{tabularx}{\columnwidth}{>{\raggedright\arraybackslash}X r r}
\toprule
\textbf{Pruning Step} & \textbf{\# of Cand.} & \textbf{Reduc. Rate} \\ 
\midrule
Original Space & $\approx 2.75 \times 10^{13}$ & - \\
+ Rule 1 & $\approx 1.14 \times 10^8$ & $>99.99\%$ \\
+ Rule 2 & $\approx 2.47 \times 10^7$ & $78.3\%$ \\
+ Rule 3 & $\approx 1.44 \times 10^7$ & $41.5\%$ \\
+ Rule 4 & $\approx 9.62 \times 10^6$ & $33.3\%$ \\
+ Rule 5 & $\approx 1.15 \times 10^6$ & $88.0\%$ \\
\midrule
\textbf{Total Reduction} & \multicolumn{2}{r}{\textbf{$>99.99\%$}} \\
\bottomrule
\end{tabularx}
\end{table}

\begin{table}[b] 
    \centering
    \small 
    \caption{Possible partitions for Spatial (S) and Temporal (T) dimensions. The letter combinations in the S and T columns are examples only.}
    \label{tab:loop_order}
    \renewcommand{\arraystretch}{0.9} 
    
    \begin{tabularx}{\columnwidth}{c c c X} 
        \toprule
        \textbf{\makecell{Num of dim \\ in S}} & \textbf{S} (Spatial) & \textbf{T} (Temporal) & \textbf{\makecell{Num of \\ schedules}} \\
        \midrule
        $1$ & \textit{M} & \textit{NKL} & $(C_4^1 \times 3! = 24)$ \\
        $2$ & \textit{MN} & \textit{KL} & $(C_4^2 \times 2! = 12)$ \\
        $3$ & \textit{MNK} & \textit{L} & $(C_4^3 \times 1! = 4)$ \\
        $4$ & \textit{MNKL} & $\emptyset$ & $(C_4^4 \times 0! = 1)$ \\
        \bottomrule
    \end{tabularx}
\end{table}


While prior work has established pruning principles for kernel fusion, these do not address the vast search space introduced by clusters and are thus insufficient for our needs. Building upon these foundations, we propose the following pruning strategies:

\begin{itemize}

    \item \textbf{Initial Search Space:} We construct our initial search space starting from the loop schedule and tile size. Drawing from methodologies in existing work, the minimum block size is set to that of a single MMA operation, i.e., $16 \times 16 \times 16$. The cluster dimension can be chosen from one of five values $\{1, 2, 4, 8, 16\}$. Since there are 4 independent dimensions, this results in $5^4$ possibilities for the cluster configuration. For a model like GPT-6.7B, we consider a problem size with $M=256, N=16384$, and $K=T=4096$. The number of valid tile choices is thus $(256/16) \times (16384/16) \times (4096/16) \times (4096/16)$. As shown in Table~\ref{tab:loop_order}, there are a total of $(24+12+4+1) = 41$ possible combinations for spatial and temporal partitioning. Therefore, the initial search space contains $(24+12+4+1) \times 5^4 \times (256/16) \times (16384/16) \times (4096/16) \times (4096/16) \approx 2.75 \times 10^{13}$ possibilities. 

    \item \textbf{Rule 1, Divisible Tile Sizes:} This is a pruning strategy from prior work~\cite{zhang2024mcfuser}, which mandates that the selected tile sizes should be hardware-aware and the problem size dimensions are evenly divisible by them. 

    \item \textbf{Rule 2, Cluster Size Constraint:} The product of cluster dimensions for each GEMM across M, N, and K must be less than the hardware limit (for H100, it is 16), and the cluster dimensions of consecutive GEMMs must be identical to ensure feasibility.  
    \item \textbf{Rule 3, Activation constraint:} To ensure the correctness of the activation between consecutive GEMMs, the accumulation dimension of preceding GEMM must be placed in the innermost loop. Otherwise, partial sums would be computed, which cannot be used by the activation and would lead to incorrect results in the subsequent GEMM. 
    \item \textbf{Rule 4, Dependency constraint:} If L dimension is set as spatial, given the dependency of GEMM, all spatial tile in L dimension will need intermediate tensor of C, but different tiles can not communicate with each other directly, therefore the fusion will fail.
    \item \textbf{Rule 5, Memory Capacity Limit:} A tensor cannot exceed the capacity of the lowest-level cache to which it can spill. 
\end{itemize} 

Among the rules above, only Rule 1 is derived from prior work\cite{zhang2024mcfuser}; the rest are novel strategies specific to this paper for handling the search space introduced by clusters. Following the analysis of prior work, the pruned search space has 11,550 ($\sim10^4$) possibilities. In contrast, our work, which considers the use of clusters, addresses a much larger search space. Therefore, a cost model is required for further analysis.






\begin{algorithm}[t]
\caption{Fusion Search Algorithm}
\label{alg:search_engine}
\begin{spacing}{0.9}
\SetKwInOut{Input}{Input}
\SetKwInOut{Output}{Output}

\Input{Graph $g$, Device $d$, Top-k count $k$}
\Output{The best execution plan $p_{best}$}
\vspace{0.5em}

\Fn{\text{SearchEngine}(g, d, k)}{
    $all\_candidates \leftarrow \text{EnumerateAllCandidates}(g, d)$\;
    $pruned\_candidates \leftarrow \text{PruneCandidates}(all\_candidates)$\;
    \vspace{0.5em}
    
    $top\_k\_list \leftarrow []$;

    \ForEach{$(s, t, r)$ in $pruned\_candidates$}{
        $(D_{v}, plan) \leftarrow \text{DataflowAnalyzer}(g, d, s, t, r)$\;
        $est\_cost \leftarrow \text{CalculateCost}(D_{v})$\;
        
        $top\_k\_list \leftarrow \text{UpdateTopKList}(top\_k\_list, (est\_cost, plan), k)$\;
    }
    \vspace{0.5em}
    
    $p_{best} \leftarrow \text{ProfileBestFromList}(top\_k\_list, d)$\;
    
    \Return{$p_{best}$}\;
}
\end{spacing}
\end{algorithm}

\subsubsection{Search Algorithm}


Algorithm~\ref{alg:search_engine} details our fusion search method. This algorithm takes a DNN graph~\texttt{g}, device information~\texttt{d}, and the top-k count~\texttt{k} as input. We first employ the pruning strategies mentioned in the previous section to filter the search space (line 3). Then, the legal candidates are fed into the DataflowAnalyzer for detailed analysis. As depicted in Algorithm~\ref{alg:spilling_planner}, we analyze and obtain the specific dataflow details under the current parameters, namely the placement of each reused tensor within the cache hierarchy and the concrete data movement volume (line 5-6). Subsequently, using the cost model described in Section~\ref{ssec:cost_model}, we iteratively evaluate each configuration to maintain a list of top-k candidates. Finally, these candidates are profiled on hardware to determine the ultimate execution plan (line 7-9). This entire search is performed offline; at runtime, kernel selection is achieved by using binning and table look-ups for the varying M dimension to select from our pre-compiled kernels. This is efficient because in FFN/conv scenarios, only the M dimension varies dynamically while N, K, and L are fixed.

\section{Implementation} \label{sec:impl}

\proj{} is a code generation framework built upon NVIDIA CUTLASS \cite{Thakkar_CUTLASS_2023}. It takes a high-level DNN model description as input and utilizes our three core components—the Fusion Search Engine, Dataflow Analyzer, and \texttt{dsm\_comm} primitive—to generate high-performance fused kernels, separating the implementation into a front-end for search and a back-end for code generation.

\subsection{Front-End: The Fusion Search Engine}
Our front-end is a Python-based search engine that explores the space of \texttt{LoopSchedules}, \texttt{TilingSizes} and \texttt{ResourceMapping}(with DSM, the lowest-level cache, selected by default). For each configuration, it invokes our Dataflow Analyzer to heuristically determine the memory mapping for intermediate results and compute the data movement volume. It then uses a cost model and pruning rules to filter candidates. The back-end is subsequently invoked to generate code. Finally, the top-$K$ configurations are passed to the hardware for on-device measurement to identify the fused kernel with the optimal performance.
\subsection{Back-End: Code Generation and Primitive Implementation}
The back-end translates the optimal plan from the front-end into high-performance CUDA code, leveraging the highly-optimized components of CUTLASS.

\paragraph{Realizing the Dataflow Analyzer}
Our heuristic plan is realized during code generation. The decision between register and smem is made by calculating the theoretical register usage for a given tile size to avoid performance-degrading spills to global memory. If SMEM is still not large enough, the data must be placed in DSM.

\paragraph{Implementing the \texttt{dsm\_comm} Primitive}
We implemented SHUFFLE, MUL, and REDUCE operations for the \texttt{dsm\_comm} primitive using a fine-grained data exchange mechanism built on TMA for data movement and the \texttt{mbarrier} intrinsic for many-to-many synchronization. Unlike the native all-to-one \texttt{cluster-sync} in CUTLASS, our \texttt{mbarrier}-based approach allows us to synchronize only the necessary groups of CTAs for a given exchange, enabling the construction of higher-level collectives like ring communication for SHUFFLE.

\paragraph{Integrating Primitives into Kernel}

Our code generator extends the CUTLASS kernel structure—prologue, mainloop, and epilogue—to orchestrate the cluster-level dataflow prescribed by the front-end. In the prologue, semaphore initialization is extended to the DSM to prepare it for inter-CTA communication. The mainloop is augmented with our \texttt{dsm\_comm} operations. For instance, upon completion of the producer's accumulation loop, a DSM \texttt{mul} is performed for GatedFFN variants to exchange and apply computation. Within the consumer's accumulation loop, a DSM \texttt{shuffle} implements a ring communication pattern to exchange intermediate results among CTAs. Finally, in the epilogue, a DSM \texttt{reduce} accumulates partial sums from different CTAs using a scatter-reduce scheme before storing the final result to global memory. This design maps the problem's spatial dimensions to the grid, while the temporal dimension to the nested execution loop within the kernel's mainloop.

\section{Evaluation} \label{sec:evaluation}

\subsection{Experimental Setup}

\begin{figure*}[t]
    \centering
    \includegraphics[width=\linewidth]{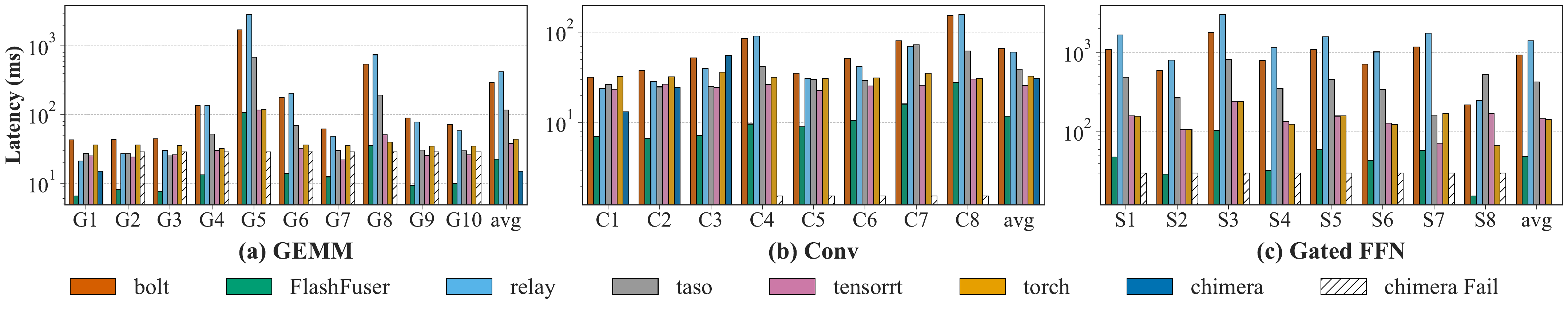}
    \caption{Performance results in various scenarios: (a) GEMM chains, (b) Convolutional chains, and (c) Gated FFNs.}
    \label{fig:eval_gemm_conv}
\end{figure*}

\begin{table}[t]
\centering
\caption{The configuration of conv chain.}
\label{tab:conv_configs}
\small 
\begin{tabular}{lccccccc}
\toprule
\textbf{ID} & \textbf{IC} & \textbf{H} & \textbf{W} & \textbf{OC1} & \textbf{OC2} & \textbf{k1} & \textbf{k2} \\
\midrule
C1 & 64 & 56 & 56 & 256 & 64 & 1 & 1 \\
C2 & 128 & 28 & 28 & 512 & 128 & 1 & 1 \\
C3 & 256 & 14 & 14 & 1024 & 256 & 1 & 1 \\
C4 & 512 & 7 & 7 & 2048 & 512 & 1 & 1 \\
C5 & 64 & 56 & 56 & 64 & 256 & 3 & 1 \\
C6 & 128 & 28 & 28 & 128 & 512 & 3 & 1 \\
C7 & 256 & 14 & 14 & 256 & 1024 & 3 & 1 \\
C8 & 512 & 7 & 7 & 512 & 2048 & 3 & 1 \\
\bottomrule
\end{tabular}
\end{table}

\begin{table}[t]
\centering
\caption{The configuration of gated FFN.}
\label{tab:gated_configs}
\small 
\begin{tabular}{cccccc}
\toprule
\textbf{ID} & \textbf{m} & \textbf{n} & \textbf{k} & \textbf{l} & \textbf{Model} \\
\midrule
S1  & 128 & 8192  & 3072 & 3072 & llama-3.2-3B \\
S2  & 128 & 5632  & 2048 & 2048 & llama-1.1B \\
S3  & 128 & 11008 & 4096 & 4096 & Llama-2-7b \\
S4  & 128 & 8192  & 2048 & 2048 & Qwen2.5-2.1B \\
S5  & 128 & 11008 & 2048 & 2048 & Qwen2.5-3B \\
S6  & 128 & 8960  & 1536 & 1536 & Qwen2.5-1.5B \\
S7  & 128 & 9728  & 2560 & 2560 & Qwen3-4B \\
S8  & 128 & 3072  & 1024 & 1024 & Qwen3-0.6B \\
\bottomrule
\end{tabular}
\end{table}

\begin{table}[t]
\centering
\caption{The configuration of gemm chain.}
\label{tab:gemm_configs}
\small 
\begin{tabular}{cccccc}
\toprule
\textbf{ID} & \textbf{m} & \textbf{n} & \textbf{k} & \textbf{l} & \textbf{Model} \\
\midrule
G1  & 128 & 512   & 32   & 256  & DLRM-0 \\
G2  & 128 & 256   & 512  & 64   & DLRM-1 \\
G3  & 128 & 512   & 416  & 256  & DLRM-2 \\
G4  & 128 & 3072  & 768  & 768  & GPT-2-Small \\
G5  & 128 & 16384 & 4096 & 4096 & GPT-6.7B \\
G6  & 128 & 4096  & 1024 & 1024 & GPT2-medium \\
G7  & 128 & 768   & 768  & 768  & nlp\_gpt3\_base \\
G8  & 128 & 8192  & 2048 & 2048 & OPT-1.3B \\
G9  & 128 & 2048  & 512  & 512  & Performer \\
G10 & 128 & 1536  & 384  & 384  & BERT \\
\bottomrule
\end{tabular}
\end{table}

\paragraph{Platforms} Our evaluation is conducted on a server-class accelerator featuring an NVIDIA H100 GPU (SXM). The host system is a dual-socket server equipped with two Intel(R) Xeon(R) Platinum 8468 CPUs (96 cores in total) clocked at 2.10GHz. The primary software stack used in our experiments includes CUDA 12.4, PyTorch 2.6, TVM 0.9, Triton 3.2, and Nsight Compute 2025.2.0.

\paragraph{Baselines} We compare \proj{} against a comprehensive set of baselines, covering industry-standard libraries and state-of-the-art research compilers.

\textbf{Libraries:} We compare against PyTorch\cite{paszke2019pytorch} 2.6 (which utilizes cuBLAS for its GEMM implementation) and NVIDIA's TensorRT\cite{NVIDIA_TensorRT_2025}, a highly optimized inference engine. For the PyTorch baseline, we enable \texttt{torch.compile}, which significantly reduces kernel launch overhead.

\textbf{Compilers:} We select several state-of-the-art machine learning compilers, including relay\cite{roesch2018relay}, TASO\cite{jia2019taso}, BOLT\cite{xing2022bolt}, and Chimera\cite{zheng2023chimera}. TVM/Relay\cite{roesch2018relay} effectively fuses kernels with a compute-activation pattern. TASO automatically performs subgraph substitutions, replacing parts of the graph with functionally equivalent but more performant alternatives (e.g., reordering consecutive matrix multiplications), but it does not support the fusion of compute-intensive operators. BOLT fuses consecutive GEMMs based on using smem and reg. Chimera implements fusion for consecutive GEMMs while also exploring different block execution orders.


\paragraph{Subgraph Configurations} The configurations of the subgraphs are detailed in Tables~\ref{tab:gemm_configs}, \ref{tab:gated_configs}, and~\ref{tab:conv_configs}. In Tables~\ref{tab:gemm_configs}\cite{touvron2023llama, hildebrand2023efficient, koroteev2021bert} and~\ref{tab:gated_configs}, the dimensions of GEMM1 are $(m \times n \times k)$ and GEMM2 are $(m \times l \times n)$. In Table~\ref{tab:conv_configs}, the dimensions are $(IC, H, W) \times (OC1, IC, K1, K1)$ for conv1 and $(OC1, H, W) \times (OC2, OC1, K2, K2)$ for conv2, where \texttt{OC1} and \texttt{OC2} are the output channel sizes of conv1 and conv2, respectively; \texttt{H} and \texttt{W} are the height and width of the feature map; and \texttt{K1} and \texttt{K2} are the respective kernel sizes.

\subsection{Subgraph Performance}
\paragraph{Performance Results} The performance evaluation results for GEMM and convolution chains are presented in Figure~\ref{fig:eval_gemm_conv}, with performance normalized to PyTorch.

\paragraph{GEMM Chains} In the GEMM chain scenario, \proj{} achieves significant speedups over all baselines, with average speedups of 5.4x over BOLT, 4.6x over Chimera, 4.7x over Relay, 3.4x over TASO, 2.4x over TensorRT, and 3.1x over PyTorch.
Although compilers like BOLT and Chimera also perform operator fusion, their methods have inherent limitations. Chimera's fusion capability is strictly limited by the SMEM size, causing it to fail on configurations with large intermediate tensors. BOLT utilizes CUTLASS templates within TVM but is constrained by its fixed block execution order, which may not be optimal. In contrast, \proj{}'s analytical model can explore a more diverse range of block execution orders. Other baselines like TASO and Relay do not fuse the two GEMMs, leading to separate kernel launches and additional global memory access overhead. Crucially, none of the above baselines leverage DSM, which fundamentally restricts their fusion scope. \proj{} overcomes these limitations by using DSM to expand the fusion boundary.

\paragraph{Convolution Chains} For convolution chains extracted from real-world ResNet models, \proj{} achieves average speedups of 6.3x over BOLT, 6.4x over Chimera, 5.6x over Relay, 4.3x over TASO, 3.3x over TensorRT, and 3.9x over PyTorch.
For smaller problem sizes, BOLT performs kernel fusion to achieve significant performance gains. However, when the problem sizes become large, BOLT abandons fusion, resulting in comparatively poorer performance. Chimera fails when convolution sizes become too large. Other baselines execute independent, non-fused convolution kernels. \proj{} utilizes DSM as a larger on-chip buffer to expand the scope of fusible operations, resulting in substantial performance gains.

\begin{figure}[htbp]
    \centering
    \includegraphics[width=\linewidth, height=0.8\textheight, keepaspectratio]{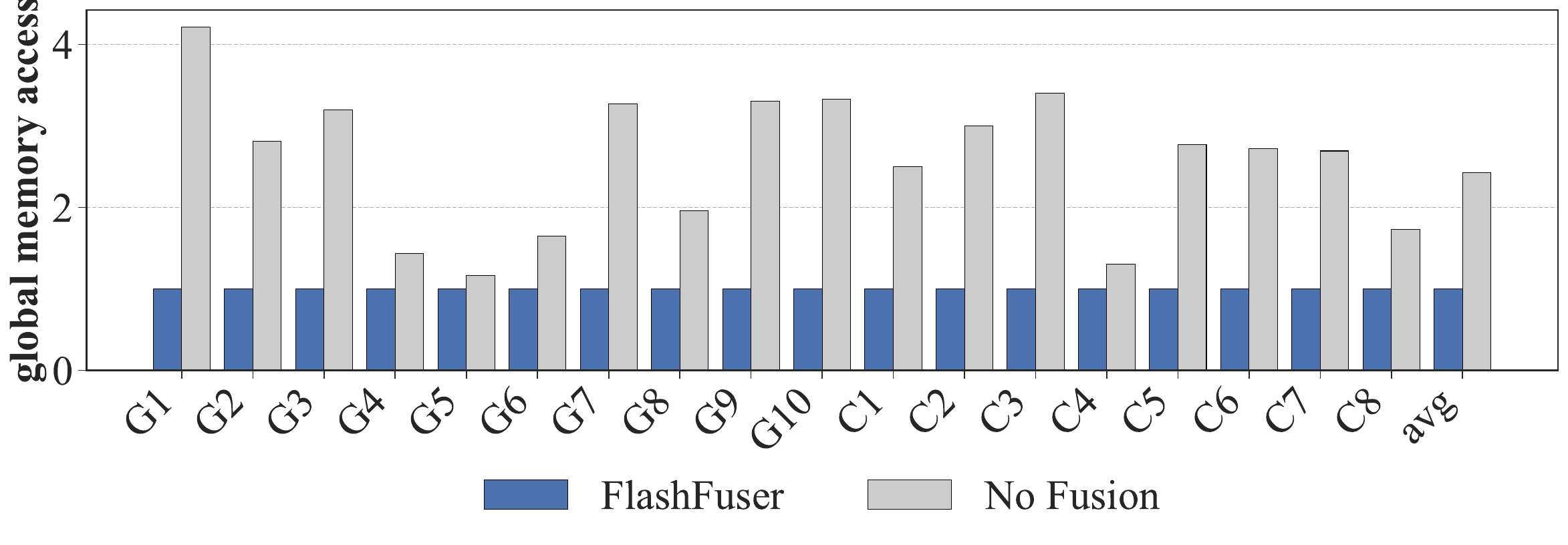}
    \caption{Comparison of global memory access between \proj{} and PyTorch.}
    \label{fig:gm_compare}
\end{figure}

\subsection{Performance Analysis}

To verify the source of the observed performance gains, we profiled the generated kernels using NVIDIA's Nsight Compute, focusing on memory access patterns. As shown in Figure~\ref{fig:gm_compare}, \proj{} significantly reduces global memory access compared to non-fused approaches like PyTorch. The analysis indicates that PyTorch, due to its lack of fusion, writes intermediate results to global memory before reading them back into shared memory for the next operator. In contrast, \proj{} enables data reuse at higher levels of the memory hierarchy, including DSM. On average, PyTorch kernels exhibit 2.4$\times$ more global memory traffic than \proj{} kernels, confirming that reduced off-chip memory access is a primary source of our acceleration.

\begin{figure}[htbp]
    \centering
    \includegraphics[width=\linewidth, height=0.8\textheight, keepaspectratio]{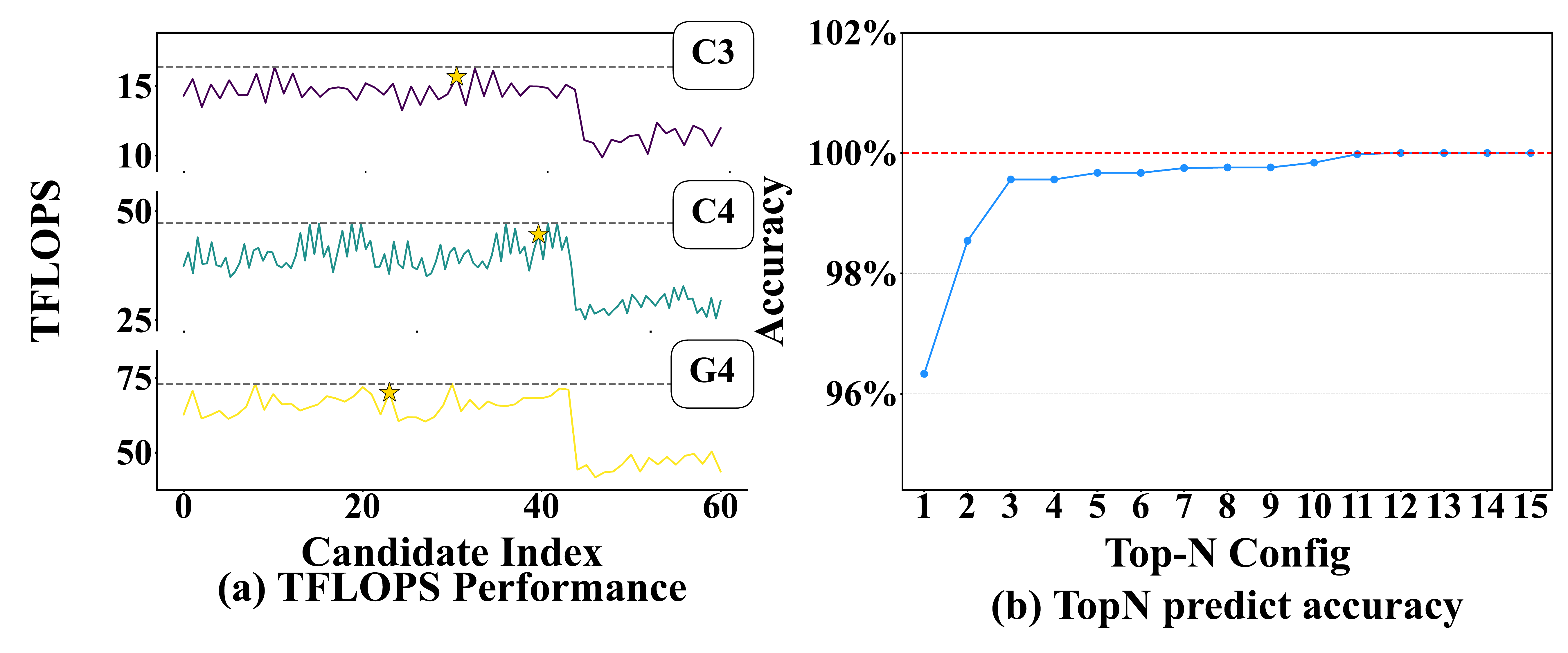}
    \caption{Validation of cost model and Analysis of top-K.}
    \label{fig:cost_model}
\end{figure}  

\begin{table}[t] 
    \centering 
    \caption{Search Time Comparison (search engine (TopK=11) vs. Brute-Force).} 
    \label{tab:search_time} 
    \begin{tabular}{@{}l c c c@{}}
    \toprule
    \textbf{} & \textbf{\makecell{Brute-Force Time}} & \textbf{\makecell{Search-Engine Time}} & \textbf{\makecell{Speedup}} \\
    \midrule
    G3 & 1.2 hr & 362.1 s & 12.25$\times$ \\
    G4 & 3.0 hr & 380.3 s & 29.05$\times$ \\
    G5 & 8.1 hr & 381.0 s & 68.26$\times$ \\
    \bottomrule
    \end{tabular}
\end{table}
To validate our cost model and search strategy, we evaluate its capability to identify optimal configurations, the selection of an appropriate \texttt{topK} value, and the compilation time overhead. Figure~\ref{fig:cost_model}a illustrates the search efficacy across the C3, C4, and G4 benchmarks. In the figure, the vertical axis represents the computing performance in TFLOPS, and different colored lines denote different models. The star markers indicate the configurations selected by our cost model. The results demonstrate that our cost model consistently identifies the performance-optimal or near-optimal configurations. Our analysis of \texttt{topK} selection (Figure.~\ref{fig:cost_model}b), using data from Table~\ref{tab:gemm_configs} and Table~\ref{tab:conv_configs}, computes accuracy as the average ratio of predicted performance to the true optimal performance. The figure shows that performance approaches 100\% as $K$ increases beyond 11, making \texttt{K=11} our chosen value. Furthermore, our search engine accelerates compilation by 12--864$\times$ compared to a brute-force search (Table~\ref{tab:search_time}), demonstrating its efficiency. This overhead primarily consists of the cost model's prediction (typically 1-2s) and the compilation time for the top-K kernels. This highlights the importance of selecting an appropriate $K$.

\begin{figure}[htbp]
    \centering
    \includegraphics[width=\linewidth, height=0.8\textheight, keepaspectratio]{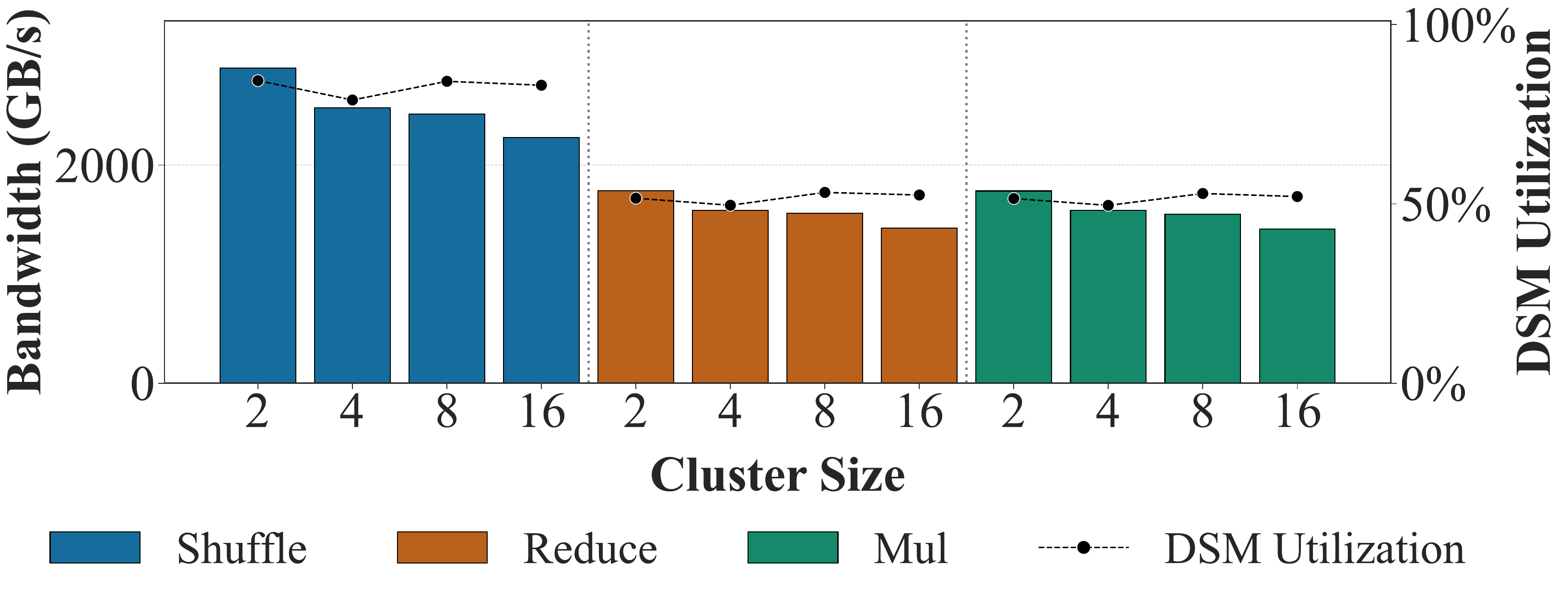}
    \caption{Bandwidth and its utilization of \texttt{dsm\_comm} primitive}
    \label{fig:dsm_prim}
\end{figure}

To validate the performance of our three proposed \texttt{dsm\_comm} primitives, we measured their bandwidth and utilization across different cluster sizes. The benchmark transfers a 32768$\times$32768 tensor, slicing it into 128x128 tiles to execute \texttt{dsm\_comm} operations within the cluster (excluding global read/store overhead), which is looped 1000 times to measure the bandwidth. Bandwidth utilization is calculated by dividing the measured bandwidth by the peak DSM bandwidth for the corresponding cluster size. As shown in Figure.~\ref{fig:dsm_prim}, while the bandwidth decreases as the cluster size increases, the bandwidth utilization remains stable. The \texttt{Shuffle} primitive outperforms \texttt{Reduce} and \texttt{Mul} because the latter two incur computational overhead in addition to data transfer.


\begin{figure}[htbp]
    \centering
    \includegraphics[width=\linewidth, height=0.8\textheight, keepaspectratio]{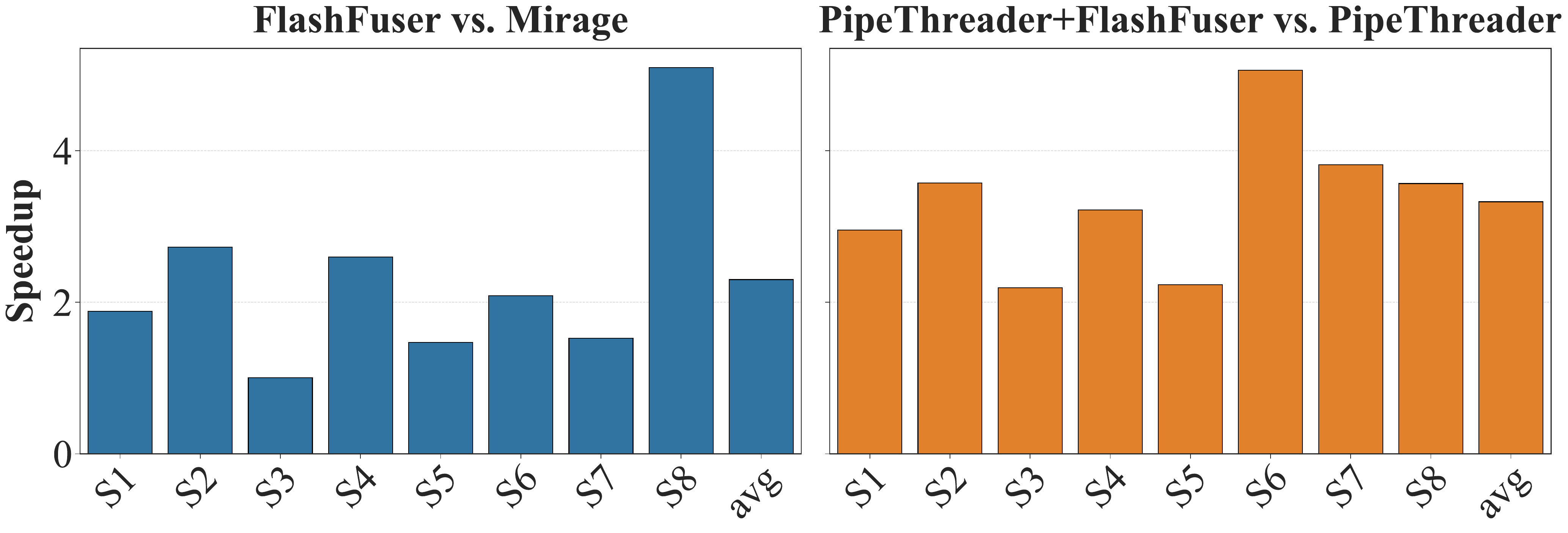}
    \caption{Comparison to mirage and pipethreader.}
    \label{fig:mirage_pipethreader}
\end{figure}
\begin{figure}[b]
    \centering
    \includegraphics[width=\linewidth, height=0.8\textheight, keepaspectratio]{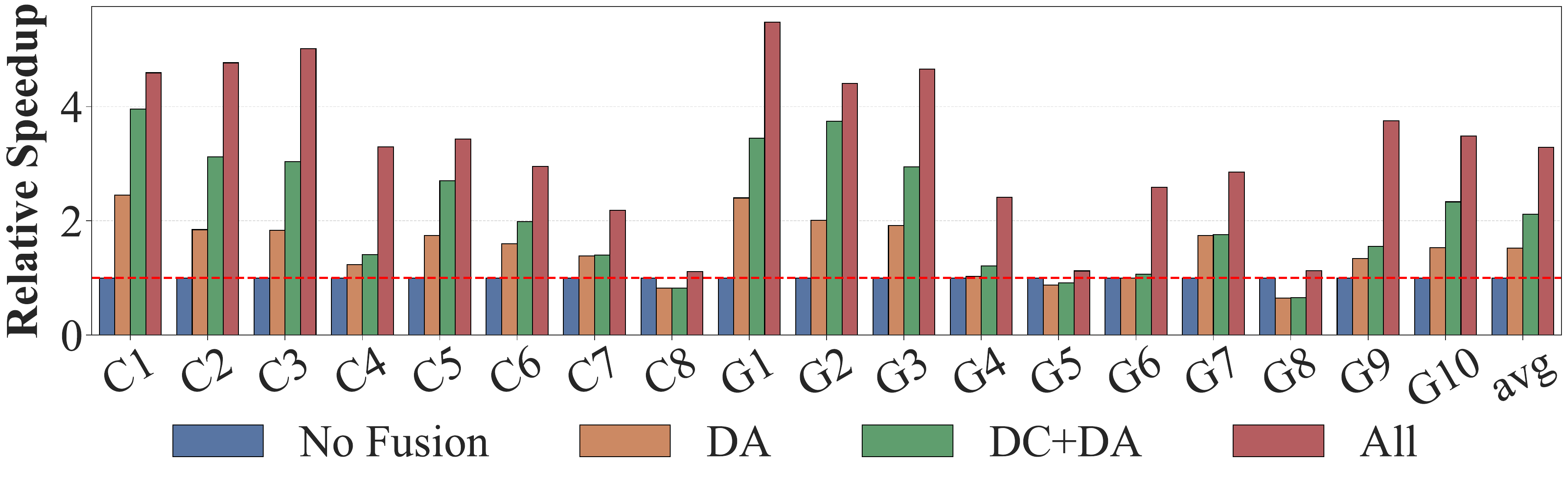}
    \caption{Ablation study of \proj{} by Isolating the Contributions of Search Engine (SE), \texttt{dsm\_comm} (DC), and Dataflow Analyzer (DA) }
    \label{fig:ablation}
\end{figure}
We conduct a detailed ablation study on our three key designs: \texttt{dsm\_comm} (DC), dataflow analyzer (DA), and search engine (SE). We evaluate the full system (`All`), `DC+DA` (using a random configuration), and `DA` (using only SMEM/global memory for fusion). As shown in Figure.~\ref{fig:ablation}, compared to a no-fusion baseline, the `All`, `DC+DA`, and `DA` configurations yield speedups of 3.29$\times$, 2.11$\times$, and 1.52$\times$, respectively. This demonstrates the effectiveness of our methods.

\begin{figure}[htbp]
    \centering
    \includegraphics[width=\linewidth, height=0.8\textheight, keepaspectratio]{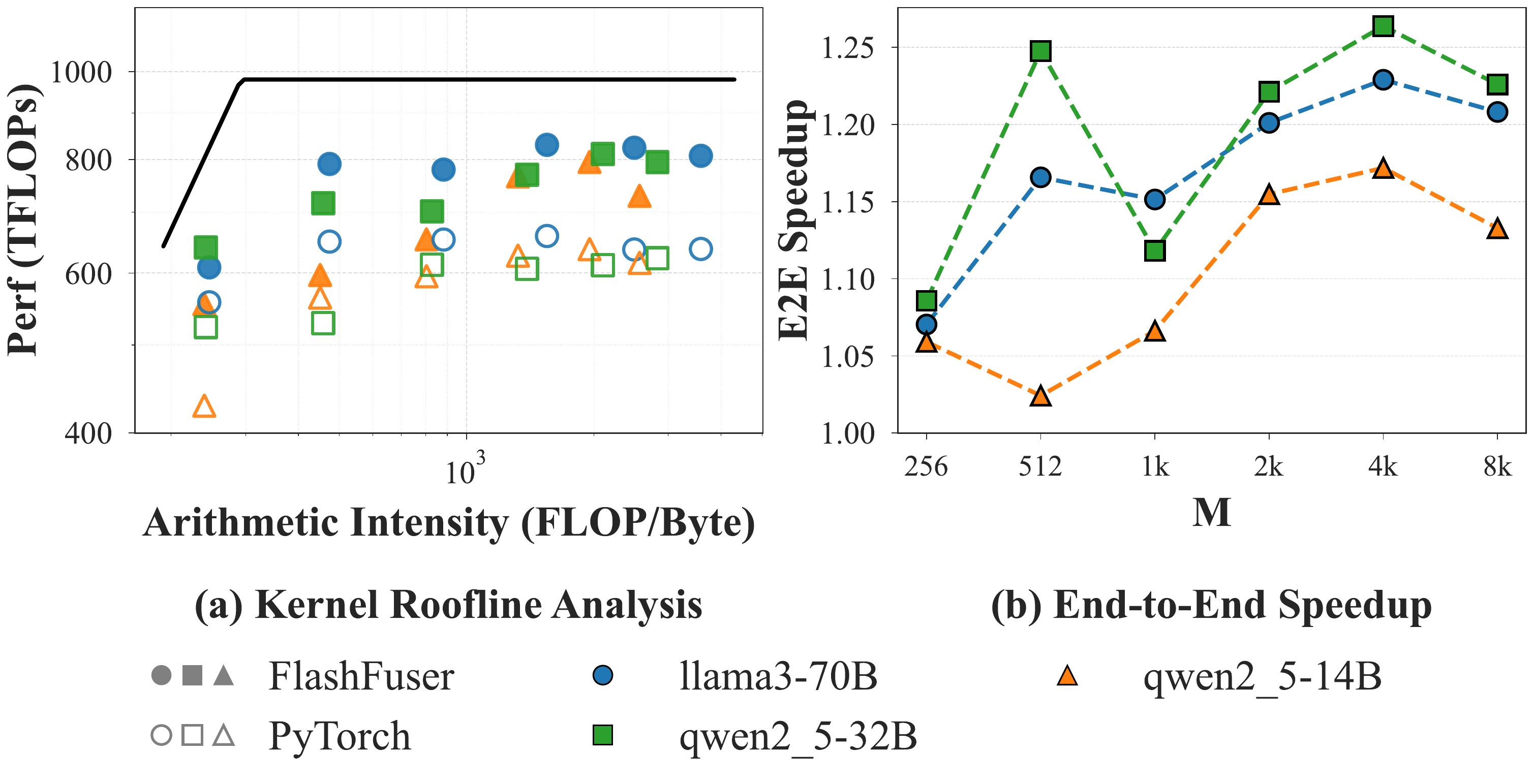}
    \caption{Kernel performance and end-to-end performance of larger LLM. }
    \label{fig:perform_large_model}
\end{figure}
\begin{figure}[t]
    \centering
    \includegraphics[width=\linewidth, height=0.8\textheight, keepaspectratio]{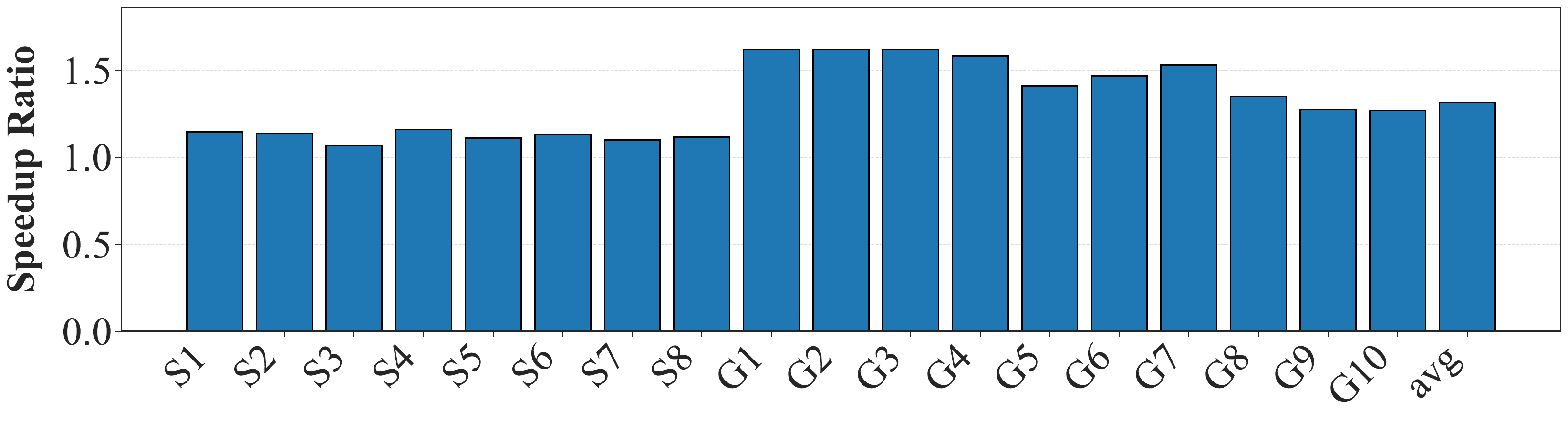}
    \caption{End-to-end performance evaluation based on SGLang.}
    \label{fig:e2e}
\end{figure}

We evaluate our end-to-end inference performance against the SGLang framework on a suite of real-world models (Table~\ref{tab:gemm_configs}/\ref{tab:gated_configs}). As illustrated in Figure.~\ref{fig:e2e}, our approach achieves an average performance improvement of 1.32$\times$. We further extend our evaluation to larger models and input sizes in Figure.~\ref{fig:perform_large_model}, testing Llama3-70B, Qwen2.5-14B and 32B. Figure~\ref{fig:perform_large_model}a presents a roofline analysis, which indicates that these models are primarily compute-bound, thus offering limited room for kernel-level optimization. In Figure~\ref{fig:perform_large_model}b, we showcase the E2E speedup. For this setup, we fix the sequence length at 256 and change batch size from 1 to 32. Across these configurations, our kernel achieves an average performance improvement of 1.22$\times$, leading to an average E2E speedup of 1.16$\times$. When considering all scenarios, including both small and large inputs, the overall E2E speedup reaches 1.24$\times$.

While our evaluation is conducted on the NVIDIA H100, the proposed fusion strategy is not limited to a specific architecture. 
FlashFuser's core abstraction, \texttt{dsm\_comm}, is a topology-agnostic collective communication concept at the design level. 
At the implementation level, for architectures with crossbar interconnects (e.g., Graphcore IPU\cite{knowles2021graphcore}, H100), our approach is directly applicable. 
For mesh architectures (e.g., Cerebras WSE\cite{matsuzaki2024performance}), a potential mapping distributes shuffle groups (defined in \S\ref{ssec:dsm_comm_primitive}) to neighboring cores to perform shuffle and reduce operations.

\section{Related Work}

While extensive research exists in both kernel fusion and Distributed Shared Memory (DSM), the intersection of these fields--how to perform efficient, automated kernel fusion on modern GPUs with DSM--remains largely explored.

\subsection{Research on Kernel Fusion}
The development of kernel fusion~\cite{zheng2023tileflow, zhuang2024mononn, xia2024optimizing}, a key compiler optimization, can be broadly categorized by the types of operators being fused.

The first primary category of fusion pairs a compute-intensive operator with subsequent memory-intensive consumers (e.g., activations, bias additions). \textit{Halide}~\cite{ragan2013halide} pioneered this for image processing pipelines with powerful schedule primitives, although for operators less complex than typical GEMMs or convolutions. Modern compilers like \textit{TVM}~\cite{chen2018tvm} and \textit{Ansor}~\cite{zheng2020ansor} advanced this by transforming loop nests to keep intermediate data in registers. To further expand the fusion scope, works like \textit{Fusion Stitching}~\cite{zheng2020fusionstitching} and \textit{AStitch}~\cite{zheng2022astitch} used shared memory as an intermediate buffer to fuse operators.

Another category is the fusion of compute-intensive operator chains (e.g., GEMM $\rightarrow$ GEMM). \textit{BOLT}~\cite{xing2022bolt} matches common patterns and invokes optimized \textit{Cutlass}~\cite{Thakkar_CUTLASS_2023} templates, though it is limited by the fixed loop schedules of Cutlass. More general transformation-based approaches include \textit{TASO}~\cite{jia2019taso}, which employs graph substitution to combine convolutions that can run in parallel, yet it lacks the capability to fuse sequential convolutions, and \textit{Chimera}~\cite{zheng2023chimera}, which optimizes at a finer grain by rescheduling dataflow between thread blocks to maximize locality.

However, a common limitation across all these works is their confinement to the resources of a single SM. This reliance forces fusion to fail when intermediate results exceed \texttt{smem}'s limited capacity. To overcome this problem, emerging hardware features like DSM have been introduced to expand the on-chip memory space.

\subsection{Research on DSM}
The study of DSM has gained traction in recent years. Researchers have explored how to design and utilize its features through various approaches, including architectural simulations and performance studies on specialized hardware.

Some research focuses on architectural exploration through simulation, proposing novel mechanisms for inter-core data sharing. For instance, Ibrahim et al.~\cite{ibrahim2020analyzing} proposed a ``shared L1'' organization to reduce redundant data replication on different L1 caches and analyzed which applications benefit from this data sharing. Falahati et al.~\cite{falahati2024cross} also interconnected L1 caches and used a predictor to determine if a cache block exists in another SM.

Other studies involve performance explorations on physical hardware that incorporates DSM. The Graphcore IPU, targeted by \textit{T10}~\cite{liu2024scaling}, has a GPU-like crossbar \texttt{smem} interconnection but assumes no HBM, a key difference from modern GPUs. The Cerebras processor, targeted by \textit{WaferLLM}~\cite{he2025waferllm}, uses a mesh interconnect L1 cache, which differs from standard GPU topology. Thus, these works have two limitations: their conclusions are not directly transferable to mainstream GPUs, and they typically focus on single-operator scheduling scenarios. Additionally, \textit{ClusterFusion}~\cite{luo2025clusterfusion} explores utilizing DSM for kernel fusion on GPUs; however, it focuses on hand-written kernels and lacks a compiler-based method for parameter selection and code generation.

While these studies highlight the potential of inter-core data sharing, a systematic compilation framework for modern GPUs is still missing. Interestingly, the concept of leveraging inter-core connections for dataflow—relatively new to general-purpose GPUs—has long been a foundational design principle in domain-specific spatial architectures.

\subsection{Fusion on Spatial Architectures}
Research on kernel fusion for specialized spatial architectures (e.g., ASIC accelerators and systolic arrays) primarily focuses on leveraging explicit on-chip Networks-on-Chip (NoC) between Processing Elements (PEs) to construct efficient dataflows. \textit{FLAT}~\cite{kao2023flat} targets memory bottlenecks in Transformer models by proposing a ``Fixed-Loop-Aligning Tiling'' strategy. It utilizes direct data reuse between PEs in a spatial array to stage intermediate results in on-chip buffers, thereby fusing originally discrete operators into a pipelined execution. \textit{COMET}~\cite{negi2025comet} introduces primitives containing explicit collectives to formally model the dataflow of compound operations, supporting the mapping of complex fusion patterns. Additionally, \textit{DESA}~\cite{wang2025desa} designs a dataflow-efficient systolic array that achieves fully fused attention computation by decoupling computation from data transfer. While these works demonstrate the efficacy of spatial dataflow, they typically rely on specific hardware interconnect topologies or systolic array structures. In contrast, \proj{} targets on GPU. It exploits the emerging DSM mechanism on modern GPUs (e.g., NVIDIA H100) to enable direct inter-core communication.

\subsection{Emerging GPU Compilers and DSLs}
To facilitate efficient code generation and optimize dataflow on GPUs, extensive research has been dedicated to machine learning compilation and Domain-Specific Languages (DSLs)~\cite{ma2020rammer,zhang2023cocktailer,zheng2020flextensor,wu2023autotuning,feng2023tensorir,zhu2022roller,zheng2022astitch,zhai2023tlp,weng2021unit,phothilimthana2019swizzle,zheng2022amos}.

Notably, \textit{Triton}~\cite{tillet2019triton} and its derivatives simplify high-performance kernel development through a block-based programming model and have been widely adopted for operator fusion. The recently proposed \textit{TileLang}~\cite{wang2025tilelang} (and its underlying low-precision library \textit{Ladder}~\cite{wang2024ladder}) advances this direction by proposing a composable tiled language and hardware-aware tensor transformations. These tools allow developers to explicitly define parallel tiling strategies and pipeline schedules across multiple memory levels via a Python interface. Although these DSLs offer powerful representation capabilities, they primarily focus on the traditional memory hierarchy and often rely on expert users to manually specify scheduling strategies. \proj{} distinguishes itself by integrating DSM into the compiler's automated search space. 
\section{Conclusion}

\label{sec:conclusion}

In this paper, we presented \proj{}, the first compiler framework that overcomes this limitation by leveraging the inter-core connection capabilities of modern GPUs. By introducing a DSM communication abstraction, using a dataflow analyzer to evaluate data placement and costs, and leveraging an efficient search engine to explore the vast search space, \proj{} systematically generates highly efficient fused kernels. On an NVIDIA H100 GPU, our evaluation shows that \proj{} delivers kernel speedups of up to $3.3\times$ against highly-tuned libraries and $4.1\times$ against state-of-the-art compilers. These gains, driven by a 58\% reduction in memory access, lead to a 1.24$\times$ end-to-end speedup.

\section*{Acknowledgment}
We thank Dr. Size Zheng for providing the source code of Chimera. This work was supported by the National Key R\&D Program of China under Grant 2022YFB4501400, and the National Natural Science Foundation of China (NSFC) Grants (62222210 and 62532006) and Shanghai Qi Zhi Institute Innovation Program SQZ202316. Any opinions, findings, and conclusions in this paper are those of the authors only and do not necessarily reflect the views of our sponsors.

\balance
\bibliographystyle{IEEEtranS}
\bibliography{refs}


\end{document}